\documentclass[prc,twocolumn]{revtex4}
\usepackage{graphicx}
\usepackage{hyperref}

\begin{document}

\newcommand{\ie}{{\it i.e.}}
\newcommand{\eg}{{\it e.g.}}
\newcommand{\etc}{{\it etc.}}
\newcommand{\cf}{{\it cf.}}
\newcommand{\etal}{{\it et al.}}
\newcommand{\be}{\begin{eqnarray}}
\newcommand{\ee}{\end{eqnarray}}
\newcommand{\jp}{$ J/ \psi $}
\newcommand{\pp}{$ \psi^{ \prime} $}
\newcommand{\ppp}{$ \psi^{ \prime \prime } $}
\newcommand{\dd}[2]{$ #1 \overline #2 $}
\newcommand{\noi}{\noindent}

\title{Tetraquarks from Intrinsic Heavy Quarks} 

\author{R. Vogt}
\affiliation
{Nuclear and Chemical Sciences Division,
Lawrence Livermore National Laboratory, Livermore, CA 94551,
USA}
\affiliation
    {Department of Physics and Astronomy,
University of California, Davis, CA 95616,
USA}

\begin{abstract}
  A number of new tetraquark candidate states containing from one to four charm or anti-charm quarks have been observed recently.  Many of these new states have been discovered at the LHC.  The production of these states via intrinsic charm in the proton is investigated.  The tetraquark masses obtained in this approach, while dependent on the internal transverse momenta of the partons in the state, are shown to agree well with the measured masses.  These calculations can provide some insight into the nature of the tetraquark candidates, whether as a bound meson pair or as a looser configuration of four individual partons.  The kinematic distributions of these states as a function of rapidity and transverse momentum are also studied.  The possible cross sections for these states are finally considered, with a comparison to the $X(3872)$ $p_T$ distributions from $p+p$ collisions at $\sqrt{s} = 13$~TeV.
\end{abstract}

\maketitle

\section{Introduction}
\label{intro}

Tetraquarks are exotic mesons that are beyond the scope of the conventional
quark model of hadrons since they contain four valence quarks.  They are denoted
as mesons because they include an equal number of quarks and antiquarks.  Such
states are not ruled out by the Standard Model but have only recently been
measured in any quantity.  The first tetraquark candidate to be reported, the
$X(3872)$, was measured by the Belle Collaboration in 2003 \cite{Belle:2003nnu}.
Later, several other states were discovered by the Belle and BESIII Collaborations, including the $Z(3900)$ \cite{Ablikim:2013,Liu:2013}. These first confirmed
measurements of tetraquark states were made at $e^+ e^-$ colliders. Although
several other states were reported elsewhere but not all have been confirmed by further
analysis \cite{SELEX:2004drx,Aaltonen:2009,D0:2016mwd,D0:2017qqm}.  Since the
advent of the LHC, many new tetraquark candidates have been discovered but still require confirmation in other measurements. 
A partial list, based on Ref.~\cite{LHCpage} is given in Table~\ref{tetraquark_table}.

\begin{table}[htbp]
  \begin{center}
    \begin{tabular}{|c|c|c|c|c|} \hline
      State &  Mass (MeV) & Quark Content & Reference  \\ \hline
      \multicolumn{4}{|c|}{states with 4 charm quarks} \\ \hline
$T_{\psi \psi}(6600)$ & $6630 \pm 90$ & $c \overline c c \overline c$ & \cite{ATLAS:2023bft} \\
  & $6552 \pm 16$ &     & \cite{CMS:2023owd} \\ \hline
$T_{\psi \psi}(6900)$ & $6905 \pm 13$ & $c \overline c c \overline c$ & \cite{LHCb:2020bwg} \\ \hline
      \multicolumn{4}{|c|}{states with 2 charm quarks} \\ \hline
$X(3872)$ & $3872 \pm 0.6$ & $c \overline u \overline c u$ & \cite{Belle:2003nnu} \\ \hline
$X_s(3960)$ & $3955 \pm 13$ & $c \overline s \overline c s$ & \cite{LHCb:2022aki} \\
      $X_s(4274)$ & $4273  ^{+10}_{-9}$ &    & \cite{LHCb:2016axx} \\
      $X_s(4500)$ & $4506 ^{+16}_{-19}$ &    & \cite{LHCb:2016axx} \\
      $X_s(4630)$ & $4630 ^{+20}_{-110}$ &   & \cite{LHCb:2021uow} \\
      $X_s(4685)$ & $4684 ^{+15}_{-17}$ &    & \cite{LHCb:2021uow} \\
      $X_s(4700)$ & $4704 ^{+17}_{-26}$ &    & \cite{LHCb:2016axx} \\ \hline
      $T_{cc}^+(3876)$ & $3870 \pm 0.12$ & $cc \overline u \overline d$ & \cite{LHCb:2021vvq} \\ \hline
      $T_{c \overline c s 1}^\theta(4000)$ & $3991 ^{+14}_{-20}$ & $c \overline c d \overline s$ & \cite{LHCb:2023hxg} \\
      $T_{c \overline c s 1}^+(4220)$ & $4220 ^{+50}_{-40}$ & $c \overline c u \overline s$ & \cite{LHCb:2021uow} \\ \hline
      \multicolumn{4}{|c|}{states with 1 charm quark} \\ \hline
      $T_{c \overline s 0}^a(2900)^0$ & $2892 \pm 21$ & $c \overline s \overline u d$ & \cite{LHCb:2022sfr} \\
      $T_{c \overline s 0}^a(2900)^{++}$ & $2921 \pm 25$ & $c \overline s u \overline d$ & \cite{LHCb:2022sfr} \\ \hline
      $T_{c s 0}(2900)^0$ & $2866 \pm 7$ & $c \overline d s \overline u$ & \cite{LHCb:2020pxc} \\
      $T_{c s 1}(2900)^0$ & $2904 \pm 5$ &  & \cite{LHCb:2020pxc} \\ \hline
    \end{tabular}
  \end{center}
  \caption[]{Some of the new particles designated as tetraquark candidate states, along with their mass and assigned quark content.  Note that the $X_s$ states, as denoted here, are often only referred to as $X$ states.  A distinction is made here for the strange quark content.  See also Ref.~\cite{LHCpage}.}
  \label{tetraquark_table}
\end{table}

Tetraquarks have long been postulated.  Indeed, when Gell-Mann proposed the idea of quarks in 1964, he also suggested that mesons  such as $q \overline q q \overline q$ and baryons such as $qqq q \overline q$ should exist as well \cite{MGM}.  In 1977 Jaffe discussed properties of $Q^2 \overline Q^2$ mesons in the context of the bag model \cite{Jaffe1,Jaffe2}.  Most models of four particle states like these containing heavy quarks have focused on $q \overline q Q \overline Q$ states and, indeed, these were the first measured, such the $X(3872)$.  The existence of $qq \overline Q \overline Q$ (and concurrently $\overline q \overline q QQ$ states) were discussed in Ref.~\cite{wiki4}.  These types of states were realized by the measurement of the $T_{cc}^+$ (and its partner the $T_{cc}^-$)  \cite{LHCb:2021vvq}.   The body of literature describing the potential structure, spectroscopy, and decays of tetraquarks is large and growing, see reviews in Refs.~\cite{Vary1,PDG_rev1,PDG_rev2,PDG,QWG_exotics} and references therein for more details.  Most treat these states as having four constituent quarks but some work also suggests that the $X(3872)$ is an excited $c \overline c$ state rather than a tetraquark \cite{Cisek:2022uqx}.  (Note also that a nuclear theory topical collaboration, the ExoHad Collaboration \cite{ExoHad}, was formed to address spectroscopy of exotic hadrons.)

Table~\ref{tetraquark_table} has been arranged according to the number of charm
quarks in the state.  A double $J/\psi$ candidate $T_{\psi \psi}(c \overline c)$
has been observed at two different mass values, 6600 MeV and 6900 MeV.  Next are
candidates with two light quarks and two charm quarks, followed by candidates
with a single charm quark, a $q \overline q q \overline Q$ configuration.  In
this latter case, one of the three light quarks is a strange quark.

Most of the tetraquark candidates listed in Table~\ref{tetraquark_table}
have quark content
$q \overline q Q \overline Q$ where $q$ or $\overline q$ can be $u$, $d$, or $s$
quarks.  The heavy quark, $Q$, can be charm or bottom.  So far, all of the
candidates reported by the LHC collaborations, see \cite{LHCpage}, contain charm
quarks.  No tetraquark candidates with bottom quarks have been measured to date.  As previously mentioned, the first $\overline q \overline q QQ$ state, the $T_{cc}^+$, was observed by LHCb, along with its antiparticle, the $T_{cc}^-$. The convention quark content given in Table~\ref{tetraquark_table} for the tetraquark candidates is defined with the $c$ quark first. However, this does not mean that the antiparticle is not detected, indeed both the particle and antiparticle are reported together.

This work aims to study the production characteristics of these tetraquark candidates, considering that they are all indeed four-quark states.  It assumes that the candidates listed in Table~\ref{tetraquark_table} will be confirmed by other measurements.  (So far only the $X(3872)$ is listed in the meson summary tables in the Review of Particle Physics \cite{PDG}.)
The calculations are performed within the intrinsic charm model,
first developed by Brodsky and collaborators \cite{intc1,intc2} in the early
1980s.  In these initial works, only the 5-particle
$|uud c \overline c \rangle$ configuration of the proton wavefunction was
considered for production of $c$ quarks, $\overline D$ and $J/\psi$ mesons and
the $\Lambda_c$ baryon.

Later work investigated double $J/\psi$ production measured by
the NA3 Collaboration \cite{Badpi,Badp} in terms of a 7-particle
$|uud c \overline c c \overline c \rangle$ configuration.  Good agreement with
the NA3 data was found \cite{dblic}.  Recently, a study of double $\Upsilon$
production from a $|uud b \overline b b \overline b \rangle$ state showed that
the double $\Upsilon$ signal reported by the A$_{\rm N}$DY Collaboration \cite{ANDY_meas} was
likely not a tetraquark candidate \cite{ANDY}.  However, one of the cases
considered in Ref.~\cite{ANDY} was compatible with the masses of the predicted
$b \overline b b \overline b$ tetraquark states
\cite{Karliner,Wang,Bai,Wu,GangYang,XZWeng}.  (It is worth noting, however that some lattice QCD calculations suggest that $b \overline b b \overline b$ tetraquarks should not be stable \cite{DAFref}.  Nonetheless, the compatibility of the calculations in Ref.~\cite{ANDY} with the previously-predicted tetraquark masses suggests that the model should be able to predict the charm tetraquark candidate masses with some reasonable accuracy.
It may also be possible to learn about the nature of the states,
whether they are tightly-bound molecules or loosely-bound four-quark configurations, by studying their mass distributions in this picture.

The intrinsic charm model is introduced and the states required to calculate the
tetraquark candidates in Table~\ref{tetraquark_table} are discussed in
Sec.~\ref{IC_prod} while the calculational structure employed to describe the tetraquark candidate sates is discussed in Sec.~\ref{CalcStruct}.  The calculations of the mass (Sec.~\ref{Sec:tet_mass}), rapidity and $p_T$
distributions (Sec.~\ref{tetra_y_pt}) are described, as well as estimates of the production cross sections (Sec.~\ref{Rates}).  Conclusions based on the results are given in Sec.~\ref{Summary}.

\section{Intrinsic Heavy Flavor Production}
\label{IC_prod}

In QCD, the wave function of a proton can be represented as a
superposition of Fock state fluctuations of the basic $\vert uud \rangle$ state,
{\it e.g.}\
$\vert uudg \rangle$, $\vert uud q \overline q \rangle$,
$\vert uud Q \overline Q \rangle\ldots$.
When the proton projectile scatters in a target, either another proton or
a nucleus, the
coherence of the Fock components is broken by a soft interaction, disrupting the state, so that the fluctuations can
hadronize \cite{intc1,intc2,BHMT}.  The model assumes that the soft interaction that disrupts the state is the only a single Fock component of the wavefunction.  No interference between Fock state configurations due to simultaneous soft interactions with two states is considered.  Such processes may exist, similar to double parton scattering in perturbative QCD but the dominant interaction there, as here, is a single interaction.  These Fock state fluctuations are
dominated by configurations with equal rapidity constituents.  Therefore,
the heavy quarks in these states carry a large
fraction of the projectile momentum \cite{intc1,intc2}.  Thus,
even though a
soft gluon interaction is all that is required to disrupt the coherence of the
state and bring the constituents on shell, the fact that the heavy quark
constituents carry a larger momentum fraction means that the can manifest themselves at forward rapidity and large $p_T$, as discussed in
Sec.~\ref{tetra_y_pt}.

While the formulation of intrinsic charm by Brodsky and collaborators is used
here, other variants of intrinsic charm distributions exist,
including meson-cloud models where the proton fluctuates into a
$\overline D(u \overline c) \Lambda_c (udc)$ state
\cite{Paiva:1996dd,Neubert:1993mb,Steffens:1999hx,Hobbs:2013bia}, also resulting
in forward production, or a
sea-like distribution \cite{Pumplin:2007wg,Nadolsky:2008zw}, only enhancing the
distributions produced by massless parton splitting functions as in DGLAP
evolution.  Intrinsic charm
has also been included in global analyses of the parton densities
\cite{Pumplin:2007wg,Nadolsky:2008zw,Dulat:2013hea,Jimenez-Delgado:2014zga,NNPDF_IC,NNPDF_IC_Nature,Guzzi1,CTEQ-TEA}.  (See Ref.~\cite{Blumlein} for a discussion of a possible
kinematic constraint on intrinsic charm in deep-inelastic scattering.)

The probability of intrinsic charm production from a 5-particle state,
$P_{{\rm ic}\, 5}^0$, obtained
from these analyses, as well as others, has been
suggested to be between 0.1\% and 1\%, see the reviews in 
Refs.~\cite{IC_rev,Stan_review} for discussions of the global analyses and other
applications of intrinsic heavy quark states.  
Evidence of a finite charm quark asymmetry in the
nucleon wavefunction from lattice gauge theory, consistent with intrinsic
charm, was presented in Ref.~\cite{Sufian:2020coz}.  

Heavy quark hadrons
can be formed from these intrinsic heavy quark states by
coalescence of the partons in the states with each other to form
either pairs of charm hadrons, such as
a $\Lambda_c^+(udc)$ and $D^0(u \overline c)$ or
a $\Sigma_c^{++}(uuc)$ and $D^-(d \overline c)$, or a $J/\psi$ and a proton
from a $|uudc \overline c \rangle$
state.  The final-state hadron is identified by its quark content and is
assumed to come on shell with its correct quantum numbers ($J^{PC}$) via a
nonperturbative process, similar to hadron production in perturbative QCD.

The $D^0$ or $D^-$ mesons
so produced have referred to as leading charm because
they can be produced from the smallest possible Fock state configuration with a
$c \overline c$ pair.  Non-leading charm, $D^+$ and $\overline D^0$, production
by intrinsic charm requires at least a 7-particle Fock state of the proton with an additional
light quark-antiquark pair, $|uudc \overline c q \overline q\rangle$.  Note that
in this case, however, $D \overline D$ pairs are produced, $D^+ D^-$ or
$D^0 \overline D^0$.  Thus these $D$ mesons would be produced with the
same momentum distributions from this state and neither is leading.
Asymmetries between
leading and nonleading charm have been measured as a function of $p_T$ and
Feynman $x$, $x_F$, or rapidity
in fixed-target $\pi^- + p$ interactions \cite{E769,E791,WA82}
and in fixed-target $p+A$ interactions
measured with the SMOG device at LHCb \cite{LHCb_leadingD}. Both scenarios
have been calculated within
the intrinsic charm picture \cite{RVSJB_asymm,RV_SMOG}.  Good agreement with the
$\pi^- + p$ data was found in Ref.~\cite{RVSJB_asymm} while the asymmetry at low
$p_T$ and near central rapidity in $p+{\rm Ne}$ interactions was underestimated
in Ref.~\cite{RV_SMOG}.

In this work, the tetraquark candidate states are assumed to be produced from the minimum possible Fock state required for their production.  In most cases, this is a 7-particle state.  For the remainder, 9-particle states are required.  Although Fock states with greater quark content could be considered, these should be less dominant, as described below.
First, higher Fock states
will have lower probabilities for production, {\it i.e.} $P_{\rm ic\, 5}^0 > P_{\rm ic\, 7}$.  Still higher Fock states will have correspondingly lower
probabilities.  Since the relative production probability is proportional to the square of the quark masses \cite{ANDY}, adding light
quark-antiquark pairs to the state results in a smaller mass penalty than
adding a heavy
$Q \overline Q$ pair, as discussed in Sec.~\ref{Rates}.
Second, as already mentioned, adding light $q \overline q$
pairs to the state results in no distinction between leading and nonleading
$D$ meson production from this state and, in fact, reduces the asymmetry when
such contributions are included \cite{tomg}.  In addition, $D$ mesons from higher
Fock states would have lower average rapidity or $x_F$ relative to the minimal
Fock state.

Because only 7- and 9-particle Fock states are employed in these calculations, in some cases, only the antiparticle of the candidate given
in Table~\ref{tetraquark_table} will be considered.  For example, production
of $T_{cc}^{+}(cc\overline u \overline d)$ would require an 11-particle Fock
state, $|uud c \overline c c \overline c u \overline u d \overline d \rangle$,
for production.  However, its antiparticle partner
$T_{cc}^{-}(\overline c \overline c u d)$ requires only a 7-particle Fock state,
$|uud c\overline c c \overline c \rangle$, for production.  An 11-particle Fock
state, $|uud c \overline c s \overline s u \overline u d \overline d \rangle$,
is required for $T_{cs}(2900)^0(c \overline d s \overline u)$ production while
its antiparticle $T_{cs}^0(\overline c d \overline s u)$ can be produced from a
7-particle state, $|uud c\overline c s \overline s \rangle$.  Thus, in thees cases, the tetraquark candidates and their antiparticles will not have identical kinematic distributions according to the model and the antiparticles will be `leading'.  On the other hand, the $T_{c \overline c s 1}(c \overline c q \overline s)$
can be produced from a 7-particle $|uud c \overline c s \overline s \rangle$ state and is thus leading while production of the antiparticle requires a 9-particle state, $|uud c \overline c s \overline s q \overline q \rangle$, for production.  %Only the leading candidate, the $T_{c \overline c s 1}$, is calculated here.

In previous work, the cross sections of $J/\psi$ and $D$ mesons from intrinsic
charm have been calculated and compared with perturbative production cross
sections \cite{RV_SMOG,RV_IC_EN}.  Though similar calculations of the cross
sections for tetraquark production from intrinsic charm states can be made,
including the appropriate additional suppression factors for production from
higher Fock states, here unit normalized distributions are shown in this
proof-of-principle calculation.  See Sec.~\ref{Rates} for estimates of the upper limits of the cross sections.

\section{Calculational Structure}
\label{CalcStruct}

The frame-independent probability distribution of a $n$-particle
Fock state in the proton containing at least one $c \overline c$ pair is
\cite{intc1,intc2}
\begin{widetext}
\be
dP_{{\rm ic}\, n} = P_{{\rm ic}\,n}^0
N_n \int dx_1 \cdots dx_n \int dk_{x\, 1} \cdots dk_{x \, n}
\int dk_{y\, 1} \cdots dk_{y \, n} 
\frac{\delta(1-\sum_{i=1}^n x_i)\delta(\sum_{i=1}^n k_{x \, i}) \delta(\sum_{i=1}^n k_{y \, i})}{(m_p^2 - \sum_{i=1}^n (m_{T \, i}^2/x_i) )^2} \, \, ,
\label{icdenom}
\ee
\end{widetext}
where $n = 7$ and 9 are the highest Fock states considered for tetraquarks
containing at least a single charm quark, as in Table~\ref{tetraquark_table}.
Here $N_n$ normalizes the probability to unity and $P_{{\rm ic}\, n}^0$
scales the unit-normalized
probability to the assumed intrinsic charm content of the proton.  The delta
functions conserve longitudinal ($z$) and transverse ($x$ and $y$) momentum.
The denominator of Eq.~(\ref{icdenom}) is
minimized when the heaviest constituents carry the largest fraction of the
longitudinal momentum,
$\langle x_c \rangle > \langle x_q \rangle, \langle x_s \rangle$.

In Table~\ref{Fock_state_table}, the tetraquark states from
Table~\ref{tetraquark_table} are shown, along with the minimal Fock state
required to produce them.  The indices 1-9 on the top row correspond to the
particle assignments in the calculations of the probability distributions
according to Eq.~(\ref{icdenom}).  To ease the calculation, the tetraquark
candidates are assigned to $i=4-7$ in an $n$-particle state.  If one imagines the
tetraquark candidate as a meson pair, one of the mesons is assigned to be the
combination of indices $4+7$ while the other is assigned indices $5+6$.
The
numerical calculations in this work based on Eq.~(\ref{icdenom}) are carried
out using the VEGAS code \cite{Vegas},
allowing up to 25 dimensional integration.

Note that the model makes no distinction between $u$ and $d$ quarks, both are assumed to have the same mass.  Therefore, the $T_{c \overline c s 1}^0(c \overline c d \overline s)$ and the $T_{c \overline c s 1}^+(c \overline c u \overline s)$ will be calculated and referred to as $T_{c \overline c s 1}$ in the remainder of this work.  Similarly, the $T_{csp}^{a \,0}(c \overline s \overline u d)$ and $T_{csp}^{a \, ++}(c \overline s  u \overline d)$ will be referred to as $T_{cs0}^a$.  In addition, the $T_{cs1}^0$ and $T_{cs0}^0$ from Table~\ref{tetraquark_table}, with the same assumed quark content ($\overline c d \overline s u)$, are listed in Table~\ref{Fock_state_table} simply as $T_{cs}^0$.

\begin{table}[htbp]
  \begin{center}
    \begin{tabular}{|c|c|c|c|c|c|c|c|c|c|c|c|} \hline
State &  Quark content & Fock state & 1 & 2 & 3 & 4 & 5 & 6 & 7 & 8 & 9 \\ \hline
$T_{\psi \psi}$ & $c \overline c c \overline c$ & $|uud c \overline c c \overline c \rangle$ & $u$ & $u$ & $d$ & $c$ & $\overline c$ & $c$ & $\overline c$ & - & -  \\ \hline \hline
$X(3872)$ & $c \overline u \overline c u$  & $|uud c \overline c c \overline c \rangle$ & $u$ & $u$ & $d$ & $c$ & $\overline c$ & $u$ & $\overline u$ & - & - \\ \hline
$X_s$ & $c \overline s \overline c s$ & $|uud c \overline c s \overline s \rangle$ & $u$ & $u$ & $d$ & $c$ & $\overline c$ & $s$ & $\overline s$ & - & - \\ \hline
$T_{cc}^-$ & $\overline c \overline c u d$ & $|uud c \overline c c \overline c \rangle$ & $u$ & $c$ & $c$ & $\overline c$ & $u$ & $\overline c$ & $d$ & - & - \\ \hline
$T_{c \overline c s}^0$ & $c \overline c  d \overline s$ & $|uud c \overline c s \overline s \rangle$ & $u$ & $u$ & $s$ & $c$ & $\overline s$ & $d$ & $\overline c$ & - & - \\ \hline
$T_{c \overline c s}^+$ & $c \overline c  u \overline s$ & $|uud c \overline c s \overline s \rangle$ & $u$ & $d$ & $s$ & $c$ & $\overline s$ & $u$ & $\overline c$ & - & - \\ \hline \hline
$T_{cs}^0$ & $\overline c d \overline s u$ & $|uud c \overline c s \overline s \rangle$ & $u$ & $c$ & $s$ & $\overline c$ & $\overline s$ & $u$ & $d$ & - & - \\ \hline
$T_{cs0}^{a\, 0}$ & $c \overline s \overline u d$ & $|uud c \overline c s \overline s u \overline u \rangle$ & $u$ & $u$ & $u$ & $c$ & $\overline u$ & $d$ & $\overline s$ & $\overline c$ & $s$ \\ \hline
$T_{cs0}^{a \, ++}$ & $c \overline s u \overline d$ & $|uud c \overline c s \overline s d \overline d \rangle$ & $u$ & $d$ & $d$ & $c$ & $\overline d$ & $u$ & $\overline s$ & $\overline c$ & $s$ \\ \hline
    \end{tabular}
  \end{center}
  \caption[]{The quark content of each type of tetraquark candidate considered
    in this work is given.  The population of the minimal Fock state required to
    product the tetraquark candidate is also given, along with the parton
    assignments used to calculate the mass distributions in
    Eqs.~(\ref{Tetra_meson_pair_mass}) and (\ref{Tetra_4quark_mass}).  Note that
    the $T_{cs1}^0$ and $T_{cs0}^0$ from Table~\ref{tetraquark_table}, with the
    same assumed quark content, are listed simply as $T_{cs}^0$.
  }
  \label{Fock_state_table}
\end{table}

In Ref.~\cite{RV_SeaQuest}, the $J/\psi$ $p_T$ distribution from intrinsic
charm was calculated for the first time by integrating over the light and charm
quark $k_T$ ranges in Eq.~(\ref{icdenom}).  In that work, $k_{T \, q}^{\rm max}$
was set to 0.2~GeV while the default for $k_{T \, c}^{\rm max}$ was taken to be
1~GeV.  The sensitivity of the results to the $k_T$ integration range was tested
by multiplying the maximum of the respective $k_T$ ranges by 0.5 and 2
respectively.  As shown in Fig.~4 of Ref.~\cite{RV_SeaQuest}, the average $p_T$
of a $J/\psi$ is slightly reduced for the smaller $k_T^{\rm max}$ but it only
narrows the $p_T$ distribution slightly, giving a final $J/\psi$ $p_T$
distribution closer to that of
a single charm quark.  On the other hand, doubling the range of $k_T^{\rm max}$
leads to a significantly wider $p_T$ distribution.  One could consider the $k_T$
range as a proxy for coherence of the bound state when calculating the mass
distributions of the tetraquark states, as discussed in Sec.~\ref{Sec:tet_mass}.

The $k_T$ range can be considered to represent the extent of the internal motion of the partons in the final-state hadron, with a low $k_T$ range corresponding to a more tightly bound state, with a lower mass and a narrower width.  A higher $k_T$ range would allow for more internal motion of the partons,, giving a wider width and a correspondingly larger size.   Thus different sets of $k_T$ integration ranges can be associated with excited states of the same meson, such as the difference between the $T_{psi psi}(6600)$ and (6900), similar to the difference in mass and radius between the $J/\psi$ and the $\psi(2S)$ in charmonium spectroscopy.

\begin{table}[htbp]
  \begin{center}
    \begin{tabular}{|c|c|c|c|c|} \hline
 Set & $k_{q}^{\rm max}$ (GeV) & $k_{s}^{\rm max}$ (GeV) & $k_{c}^{\rm max}$ (GeV) & 
$k_{\rm TQ}^{\rm max}$ (GeV)  \\ \hline 
kt1 & 0.2 & 0.4 & 1.0 & 1.0 \\
kt2 & 0.1 & 0.2 & 0.5 & 0.5 \\
kt3 & 0.4 & 0.8 & 2.0 & 2.0 \\  
kt4 & 0.3 & 0.6 & 1.5 & 1.5 \\ \hline
    \end{tabular}
  \end{center}
  \caption[]{The four sets of maximum range of $k_T$ integration for light quarks, strange quarks (when applicable), charm quarks, and the tetraquark state in the proton, designated by the subscript TQ here.}
  \label{kt_set_table}
\end{table}

Several different scenarios are considered for the transverse momentum ranges
of the quarks and constituent mesons of the tetraquark states.  They are given
in Table~\ref{kt_set_table}.  The set kt1 takes the default values used for
$J/\psi$ and $\overline D$ meson calculations in Refs.~\cite{RV_SMOG,RV_IC_EN}
while with
sets kt2 and kt3 the range is halved (for a more tightly bound state) and
doubled (for a more weakly bound state).  A fourth set, kt4, takes values 50\%
higher than those of kt1.

The quark masses used in the calculation of the transverse masses in
Eq.~(\ref{icdenom}) are $m_q = 0.01$~GeV, $m_s = 0.3$~GeV and $m_c = 1.27$~GeV,
effectively current quark masses.  
The kinematic distribution can be calculated
assuming simple coalescence of the quark in the state described in
Eq.~(\ref{icdenom}) by adding the appropriate delta functions in the $x$ and $k_T$
directions.

As shown in Ref.~\cite{RV_IC_EN}, while the $x_F$ distribution is independent of
$k_T$ and initial energy, the rapidity distribution is boosted in the 
direction of the incident proton.  One starts with the calculation of the
$x_F$ distribution and uses a Jacobean in the integral to transform to rapidity.
The $x_F$ or rapidity-integrated $p_T$ distribution does not depend on the center of mass energy of a collision bringing a tetraquark state in the proton, as in
Eq.~(\ref{icdenom}), on mass shell.  However, the $p_T$ distribution in a
specific rapidity region is
significantly modified.  The higher the energy, the more the average $p_T$ is
moved to higher $p_T$ at midrapidity.  The distribution is suppressed less at
forward rapidity in high energy collisions, see Ref.~\cite{RV_IC_EN}.  This
will be discussed in more detail in Sec.~\ref{tetra_y_pt}.

\section{Tetraquark Mass Distributions}
\label{Sec:tet_mass}

In these calculations, charm tetraquark production is first assumed to occur
only as a pair of charm mesons for states with two or more charm quarks.  In
these cases, the masses of the two meson components are well balanced and a
mass peak can be found.  This is the case for $T_{\psi \psi}$, $X(3872)$, $X_s$
and $T_{cc}^-$.  If one considers the $T_{c \overline c s}$ as a $(c \overline c)+(q \overline s)$ 
meson pair or the $T_{cs}$ states as either 
$D_s (c \overline s) + \pi (q \overline q^\prime)$ for $T_{c \overline s}$ or
$D^+ (c \overline d) + K^- (s \overline u)$ for $T_{cs}$, one does not find a
distribution with a clear maximum mass because
the light meson could be described as orbiting the
heavy charm meson with higher masses corresponding to larger distances between
the two.  However, if these states are simply described as an uncorrelated
cluster of two quarks
and two antiquarks, the state is stable with a finite mass.

Such differences in the stability of the state based on the assumed internal structure can result in different final-state properties that could be distinguished in other production processes.  For example, in heavy-ion collisions, tetraquark candidates that can exist as a pair of bound mesons may be harder to break up in the produced hot medium than those that are made up of  four independent partons.  On the other hand, one might expect that tetraquarks could also be produced by recombination of constituent partons within sufficiently close proximity in high multiplicity heavy-ion collisions, as discussed in Refs.~\cite{sRef1,sRef2}.

In this section,
the two ways of calculating the tetraquark mass distributions are described.
The calculated mass distributions from both cases are shown in Sec.~\ref{Tet_Mass_Dists}.

\subsection{Tetraquark Production as a Meson Pair}
\label{double_Ups}

If the tetraquark is considered to be a bound pair of
mesons, the mass distribution can be calculated as \cite{dblic,ANDY}
\begin{widetext}
\begin{eqnarray}
\lefteqn{\frac{dP_{{\rm ic}\, n}}{dM^2_{\rm TQ}} =
  \int \frac{dx_{M_1}}{x_{M_1}} \frac{dx_{M_2}}{x_{M_2}}
  \int dm_{M_1}^2 dm_{M_2}^2
  \int dk_{x \, M_1} dk_{y \, M_1}dk_{x \, M_2} dk_{y \, M_2}
  \int \frac{dx_{\rm TQ}}{x_{\rm TQ}}\int dk_{x \, {\rm TQ}}
  dk_{y \, {\rm TQ}}\ dP_{{\rm ic}\, n}} \nonumber \\ 
  &   & \mbox{} \times \delta \left( \frac{m^2_{T, M_1}}{x_{M_1}} -
  \frac{m_{T \, 4}^2}{x_{4}} - \frac{m_{T \, 7}^2}{x_{7}} \right)
  \delta(k_{x \, 4} + k_{x \, 7} - k_{x \, M_1})
  \delta(k_{y \, 4} + k_{y \, 7} - k_{y \, M_1}) 
  \delta(x_{M_1} - x_4 - x_{7}) \nonumber \\
  &   & \mbox{} \times \delta \left( \frac{m^2_{T, M_2}}{x_{M_2}} -
  \frac{m_{T \, 5}^2}{x_{5}} - \frac{m_{T \, 6}^2}{x_{6}} \right)
  \delta(k_{x \, 5} + k_{x \, 6} - k_{x \, M_2})
  \delta(k_{y \, 5} + k_{y \, 6} - k_{y \, M_2})
  \delta(x_{M_2} - x_5 - x_{6}) \nonumber \\
&  & \mbox{} \times\delta
  \left( \frac{M^2_{T, {\rm TQ}}}{x_{\rm TQ}} -
  \frac{m_{T, M_1}^2}{x_{M_1}} - \frac{m_{T, M_2}^2}{x_{M_2}} \right)
\delta(k_{x \, M_1} + k_{x \, M_2} - k_{x \, {\rm TQ}})
\delta(k_{y \, M_1} + k_{y \, M_2} - k_{y\, {\rm TQ}})
\delta(x_{\rm TQ} - x_{M_1} - x_{M_2})
\label{Tetra_meson_pair_mass}
\end{eqnarray}
\end{widetext}
where $dP_{{\rm ic}\, n}$ is taken from Eq.~(\ref{icdenom}).  As written here,
the pair mass distributions require integration over the invariant mass of each
meson of the pair, $M_1$ and $M_2$, over some suitable range.  In the
calculation of $T_{\psi \psi}$
in this work, as well as previous calculations of double $J/\psi$ \cite{dblic}
and $\Upsilon$ \cite{ANDY} production, the range is between twice the heavy
quark mass and twice the mass of the lowest mass heavy meson,
$2m_Q < m_{T \, M} < 2m_H$.  In the other cases, where charm mesons are
considered, a delta function,
$\delta (m^2_{M_i} - m^2_{M_{\rm meson}})$ is employed to integrate over
$dm^2_{M}$.  For example, to calculate $X(3872)$, $m_{M,1}$ are $m_{M_2}$ taken to
be $m_{D^0}$ and $m_{\overline D^0}$ respectively.
The delta functions ensure conservation of momentum for both mesons in the pair
as well as the tetraquark candidate.  The tetraquark candidate is labeled by the
subscript TQ.

The mass-dependent delta functions in Eq.~(\ref{Tetra_meson_pair_mass}) can be
considered to constitute the minimal assumption of a tetraquark wavefunction in
the intrinsic heavy quark state.  In this case, the delta functions for mesons
$M_1$ and $M_2$ represent a two-body wavefunction for each meson while that for
$M_{\rm TQ}$ represents the two-body meson-pair wavefunction.  Thus the constituents of each two-body system can have any kinematics that does not exceed the $k_T$ cutoffs as long as each system conserves momentum.  The $k_T$ ranges emphasized in Table~\ref{kt_set_table} are thus related to how tightly the mesons and the tetraquark itself are bound.

As was shown in Ref.~\cite{ANDY}, the $k_T$ integration range affects
the mass distributions.  A narrower $k_T$ integration range
resulted in a mass distribution with a lower average value and a narrower
width.  Thus specific $k_T$ ranges can differentiate between excited states.
This possibility is explored for
the $X_s(c \overline s \overline c s)$ states in this section.

Note that current quark masses are used in these calculations
because $m_{M}$ has to be larger than $\hat{m}_{T,i}$
for the delta functions describing the mass differences in
Eq.~(\ref{Tetra_meson_pair_mass}) to have a solution. 
This condition would not be satisfied if constituent quark masses are employed
for {\it e.g.} $T_{c \overline c s 1}$ as a
$(c \overline c)(q \overline s) \equiv J/\psi + K$ meson pair
because $m_q + m_s > m_K$ if
constituent quark masses are used.
Even employing these masses, the probability as a function
of tetraquark candidate mass rises with $M_{\rm TQ}$ rather than exhibiting a
resonance structure.  This is because, unlike a bound
meson pair of two nearly equal objects orbiting around each other, the physical
picture is
more like a light object orbiting around a stationary heavy object.  Higher
values of $M_{\rm TQ}$ imply a larger radius where the lighter meson orbits
further away from the heavy one.  Therefore, these systems are treated
as a cluster
of four independent quarks, as shown in the next section.

\subsection{Tetraquark Production as a Four Quark State}
\label{Xb_tet7}

Here the constraint that the four quarks in a tetraquark candidate are
considered to be part of a meson-antimeson pair is relaxed.  If the four quarks
constituting a tetraquark candidate are considered as a single cluster, the
mass distribution for $M_{\rm TQ}$ is then \cite{ANDY}
\begin{widetext}
\be
\frac{dP_{{\rm ic}\, n}}{dM^2_{\rm TQ}} & = &
  \int \frac{dx_{\rm TQ}}{x_{\rm TQ}}\int dk_{x \, {\rm TQ}}
  dk_{y \, {\rm TQ}}\ dP_{{\rm ic}\, n} \,\,
\delta\left( \frac{M^2_{T, {\rm TQ}}}{x_{\rm TQ}} - \frac{m_{T, 4}^2}{x_{4}}
- \frac{m_{T, 5}^2}{x_{5}}- \frac{m_{T, 6}^2}{x_{6}}- \frac{m_{T, 7}^2}{x_{7}}
\right) \label{Tetra_4quark_mass}
\\ 
&  & \mbox{} \times 
\delta(k_{x \, 4} + k_{x \, 5} + k_{x \, 6} + k_{x \, 7}
- k_{x \, {\rm TQ}})
\delta(k_{y \, 4} + k_{y \, 5} + k_{y \, 6} + k_{y \, 7}
- k_{y \, {\rm TQ}})
\delta(x_{\rm TQ} - x_4 - x_5 - x_6 - x_7) \, \, , \nonumber
\ee
\end{widetext}
where $dP_{{\rm ic}\, n}$ is from Eq.~(\ref{icdenom}). 
Under this assumption, the $T_{c \overline c s 1}$, $T_{c \overline s 0}^a$, and $T_{c s}^0$
mass distributions all have a discernible mass peak.

In this case the mass-dependent delta functions in
Eq.~(\ref{Tetra_meson_pair_mass}) are replaced by a single delta function
connecting the four quarks in the states independently rather than considering
them as constituents of mesons internal to the tetraquark wavefunction.
Here the $k_T$ ranges in Table~\ref{kt_set_table} set the scale for the
proximity of the individual quarks in momentum space rather than as partners in a meson.

The mass distributions
for both assumptions of the tetraquark structure,
Eqs.~(\ref{Tetra_meson_pair_mass}) and (\ref{Tetra_4quark_mass}), 
are shown in the next section.

\subsection{Calculated Mass Distributions}
\label{Tet_Mass_Dists}

This section presents the mass distributions calculated as described in the
previous subsections.  The average masses and widths will also be reported.
The average masses, calculated using both Eqs.~(\ref{Tetra_meson_pair_mass})
and (\ref{Tetra_4quark_mass}), as discussed in the preceding sections,
can be found in Table~\ref{ave_mass_table}.  The widths of the distributions are
given in 
Table~\ref{FWHM_table}.  The widths are calculated as the standard deviation of
the mass distributions.  Thus, if the mass is defined as the average over the
distribution, $\langle M_{\rm TQ} \rangle$, the variance of the distribution is
$\langle M_{\rm TQ}^2 \rangle - \langle M_{\rm TQ} \rangle^2$ and the standard
deviation is the square root of the variance.

Because the focus here is on production by coalescence, with no constraints other than the momentum-conserving delta functions, the model cannot reproduce the masses and widths of the tetraquark candidates on the few-MeV or lower scales of the measured values, but only on the order of several hundred MeV.  As will be shown, the masses are in rather good agreement with the measured values without introducing further constraints.  The widths are, unsurprisingly, considerably overestimated.

The first set of mass distributions, corresponding to states that can be
calculated under the assumption that the state is produced as a meson pair,
are shown in Sec.~\ref{Tet_Meson_Mass}.  The second set of distributions,
assuming the tetraquark is produced as a four-quark state, are shown in
Sec.~\ref{Tet_4q_Mass}.  All mass 
distributions shown have the probability normalized to unity since the focus
is on the shape of the distributions.  The possible production cross sections
are discussed in Sec.~\ref{Rates}.

\subsubsection{Meson Pair Distributions}
\label{Tet_Meson_Mass}

The mass distributions of the $T_{\psi \psi}$, with four charm quarks, 
$c \overline c c \overline c$, are shown first.  Then the $X(3872)$, the $X_s$
states, and the $T_{cc}^-$, all calculated assuming the state consists of a
$D \overline D$ or $D_s \overline D_s$ meson pair, are shown.

\begin{figure}
  \begin{center}
    \includegraphics[width=0.495\textwidth]{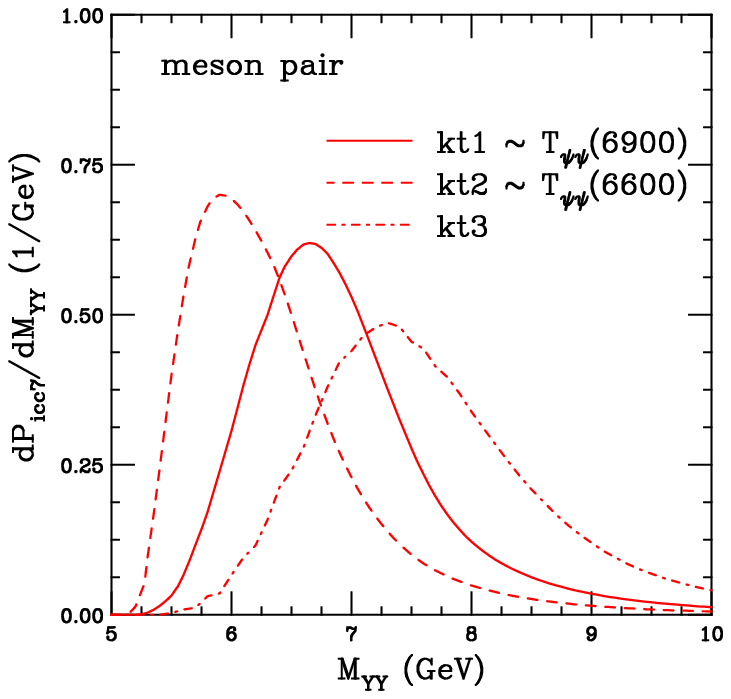}
  \end{center}
  \caption[]{The $T_{\psi \psi}$ probability distribution, calculated using
    Eq.~(\ref{Tetra_meson_pair_mass}),
    as a function of mass of the state.
    Calculations are shown for kt1 (solid), kt2 (dashed), and kt3
    (dot-dashed).
  }
\label{ic_Tpp_Mdists}
\end{figure}

Figure~\ref{ic_Tpp_Mdists} shows the results for the $T_{\psi \psi}$ for three of
the four $k_T$ ranges: kt1, kt2 and kt3.  The distributions are all rather
broad, with the distributions becoming broader as the $k_T$ integration range is
increased. There is a distinct shift to higher average masses with increasing
maximum $k_T$.  The lowest $k_T$ maximum, corresponding to kt2, gives the lowest
average mass, approximately consistent with the $T_{\psi \psi}(6600)$,
$M = 6.36$~GeV.  The $k_T$ maximum used as
the default in Ref.~\cite{ANDY}, kt1, gives an average mass consistent with 
$T_{\psi \psi}(6900)$, 6.93~GeV.  The average mass of all three cases can be
found in Table~\ref{ave_mass_table}.  The widths of the distributions are given
in  Table~\ref{FWHM_table}.  They are between 0.75 and 0.9~GeV with the
narrowest width associated with the lowest mass. 

\begin{figure}
  \begin{center}
    \includegraphics[width=0.495\textwidth]{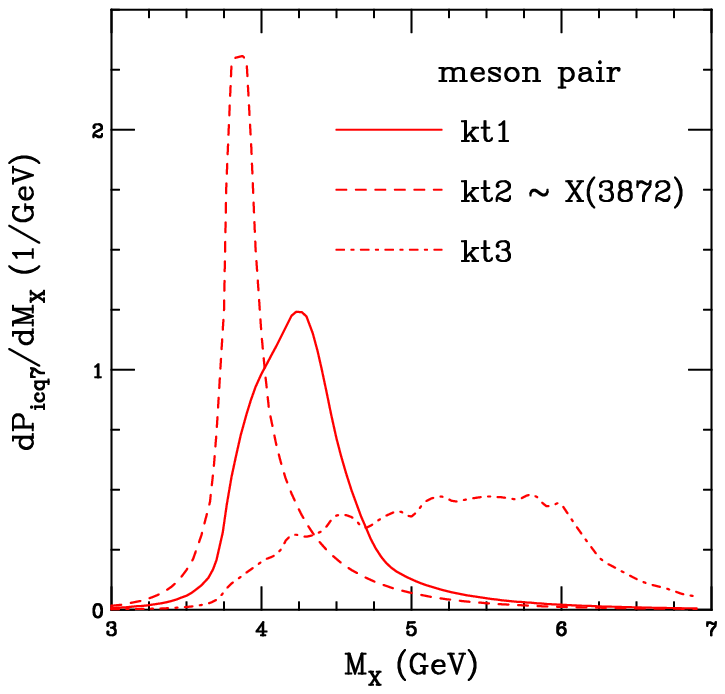}
  \end{center}
  \caption[]{The $X(3872)$ probability distribution,  calculated using
    Eq.~(\ref{Tetra_meson_pair_mass}),
    as a function of mass of the state.
    Calculations are shown for kt1 (solid), kt2 (dashed), and kt3
    (dot-dashed).
  }
\label{ic_X_Mdists}
\end{figure}

The mass distributions for the $X(3872)$ are shown in Fig.~\ref{ic_X_Mdists}.
The average mass assuming set kt1 for the $k_T$ range is 4.3~GeV, significantly
larger than the mass of the $X$ but could be in agreement with an excited state of the $X$, as seen for $T_{\psi \psi}$ in Fig.~\ref{ic_Tpp_Mdists}.  However, when the narrower range of set kt2 is
used, the average mass is 4.0~GeV, only approximately 0.1~GeV more than the
measured mass.  In addition, the width in this case is 0.42~GeV, broader than
the measured width but resulting in a
rather narrow peak nonetheless.  Doubling the
$k_T$ range, with set kt3, results in a broader distribution with a width of
$\sim 0.74$~GeV.  

Given the rather good agreement of set kt2
with the measured $X(3872)$ mass, most
of the calculations of the kinematic distributions in Sec.~\ref{tetra_y_pt}
will employ this assumption.  Exceptions will be made for the $T_{\psi \psi}$
and the $X_s$ states which have reported more than one mass state.

\begin{figure}
  \begin{center}
    \includegraphics[width=0.495\textwidth]{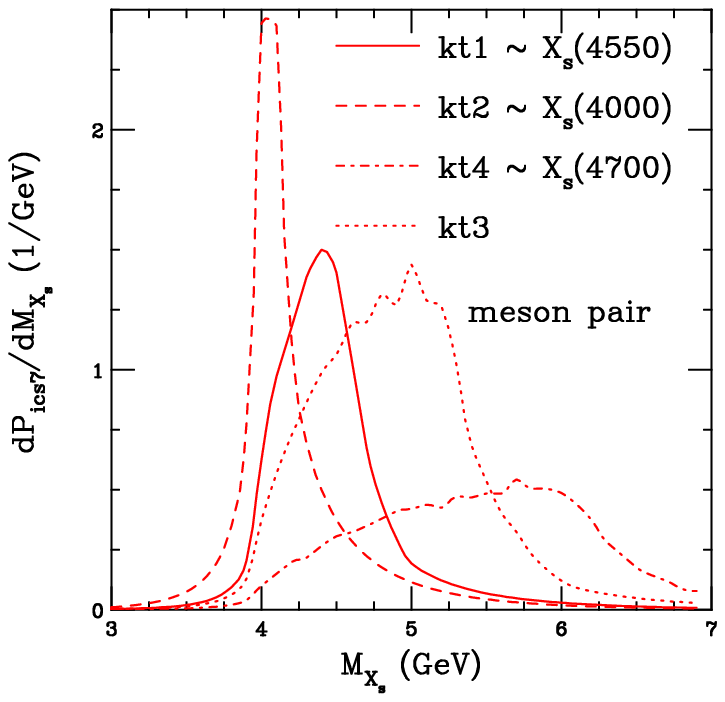}
  \end{center}
  \caption[]{The $X_s$ probability distributions,  calculated using
    Eq.~(\ref{Tetra_meson_pair_mass}),
    as a function of mass of the state.
    Calculations are shown for kt1 (solid), kt2 (dashed), kt3
    (dot-dashed) and kt4 (dotted).  The approximately grouped $X_s$
    masses based on
    Table~\ref{tetraquark_table} are associated with the closest $k_T$ range.
  }
\label{ic_Xs_Mdists}
\end{figure}

Figure~\ref{ic_Xs_Mdists} shows the mass distributions of the states labeled
$X_s$ here, with $c \overline s \overline c s$ content, effectively a tetraquark
state composed of a $D_s \overline D_s$ pair.  Based on
Table~\ref{tetraquark_table}, the measured $X_s$ states
can be approximately grouped into masses of 4.0, 4.55 and 4.7~GeV.  To better
approximate these masses, set kt4 was introduced with a $k_T$ range intermediate
between sets kt1 and kt3.  Indeed, the calculations shown in
Fig.~\ref{ic_Xs_Mdists} for sets kt2, kt1 and
kt4, with average masses of 4.2, 4.5 and 4.9~GeV respectively, are in rather
good agreement with these approximate masses.  More fine tuning of the $k_T$
range could separate individual masses further.  

The width increases slowly with the $k_T$ range.  The widths with sets kt1 and
kt2 are similar, 0.43~GeV, while the width with set kt4 is 0.53~GeV.  The
average mass and
corresponding width employing set kt2 is somewhat
higher than that for the $X(3872)$ in 
Fig.~\ref{ic_X_Mdists} because the light $q \overline q$ in the 7-particle Fock
state for the $X(3872)$ is replaced by the heavier $s \overline s$ pair.

\begin{figure}
  \begin{center}
    \includegraphics[width=0.495\textwidth]{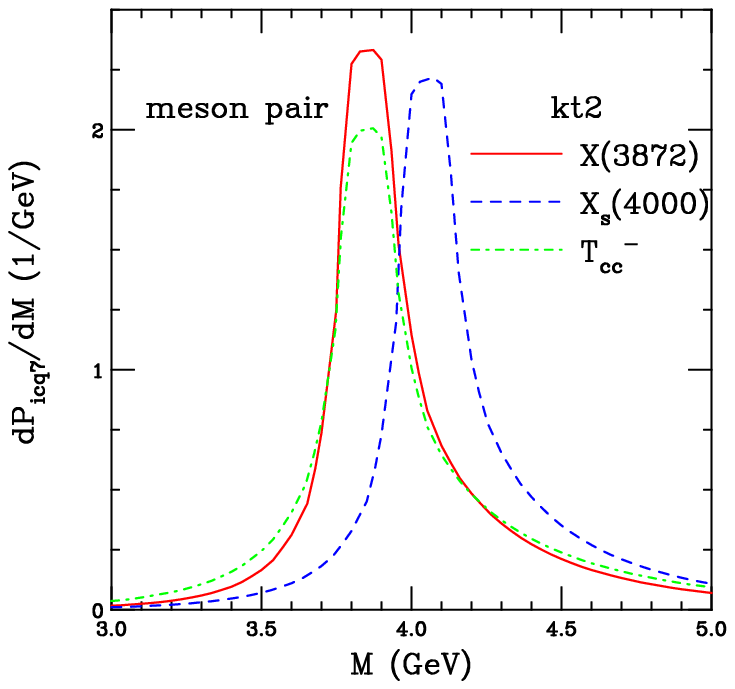}
  \end{center}
  \caption[]{The probability distributions for $X(3872)$ (red solid),
    $X_s$ (blue dashed) and $T_{cc}^-$ (dot-dashed green),  calculated using
    Eq.~(\ref{Tetra_meson_pair_mass}), as a function of mass of the state
    for scenario kt2.
  }
\label{ic_X_Xs_Tcc_Mdists}
\end{figure}

Figure~\ref{ic_X_Xs_Tcc_Mdists} compares the mass distributions of the
$X(3872)$, $X_s$ and $T_{cc}^-$ for set kt2.  The mass shift between the
$X(3872)$ and the $X_s$ with $M \approx 4.0$~GeV is apparent, as is the somewhat
broader width.  The $T_{cc}^-$
mass is almost identical to that of the $X(3872)$, as may be expected
due to the nearly identical quark content.  Its width is also similar, 0.5~MeV
compared to 0.4~MeV for the $X(3872)$.

\subsubsection{Mass Distributions Assuming a Four-Quark State}
\label{Tet_4q_Mass}

Here the mass distributions of tetraquark states assumed to be composed of
loosely bound 4-quark states are shown.  These states have either one charm
quark or, in the case of $T_{c \overline c s}$, the $c$ and $\overline c$ quarks
are assumed to form a ``$\psi$" rather than connect
to the light quarks in the
state.  If the $c$ and $\overline c$ are assumed to be associated with each
other as in a $J/\psi$, no well-defined mass can be obtained.  However, if one,
on the other hand, assumed that the parton configuration was $(c \overline q) (c \overline s) \equiv D D_s$, one could calculate this state in a meson pair
configuration.  The masses given in Table~\ref{tetraquark_table} are similar to
the two  lowest masses given for the $X_s$ tetraquark candidates.

\begin{figure}
  \begin{center}
    \includegraphics[width=0.495\textwidth]{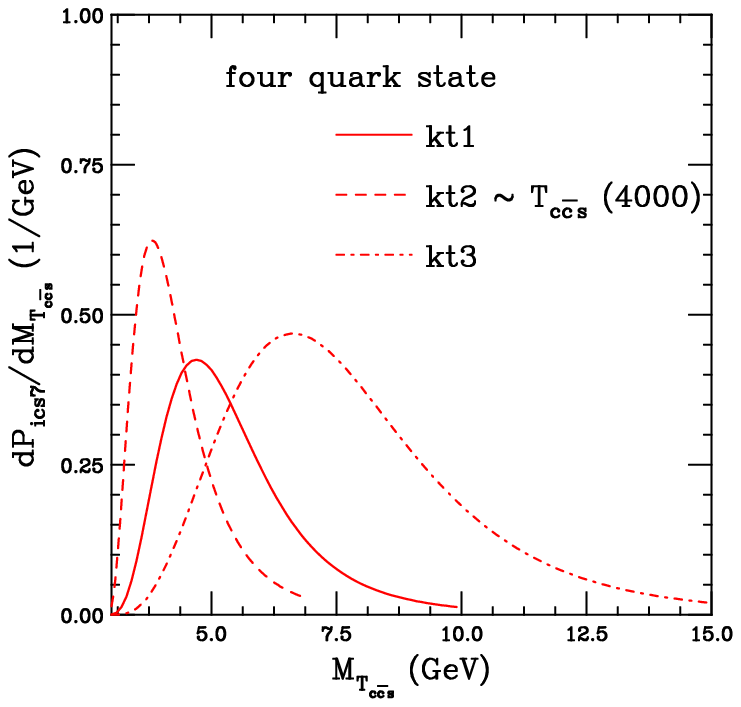}
  \end{center}
  \caption[]{The $T_{c \overline c s}$ probability distribution, calculated using
    Eq.~(\ref{Tetra_4quark_mass}),
    as a function of mass of the state.
    Calculations are shown for kt1 (solid), kt2 (dashed), and kt3
    (dot-dashed).
  }
\label{ic_Tpsis_Mdists}
\end{figure}

The $T_{c \overline c s}$ mass distributions are shown in Fig.~\ref{ic_Tpsis_Mdists}.
As in the previous section, the distributions with larger $k_T$ ranges have
larger average masses.  However, now since the quarks are not assumed to pair
into mesons but exist in a more loosely bound configuration, the widths are
considerably larger, even for the narrowest distribution
calculated with set kt2.  The average mass with this set, 4.3~GeV, is very
similar to that of the $X_s$ calculated with set kt2 but the width is now
$\sim 0.8$~GeV instead of the
value of 0.4~GeV obtained for the $X_s$ calculated with the same $k_T$ set.

\begin{figure}
  \begin{center}
    \includegraphics[width=0.495\textwidth]{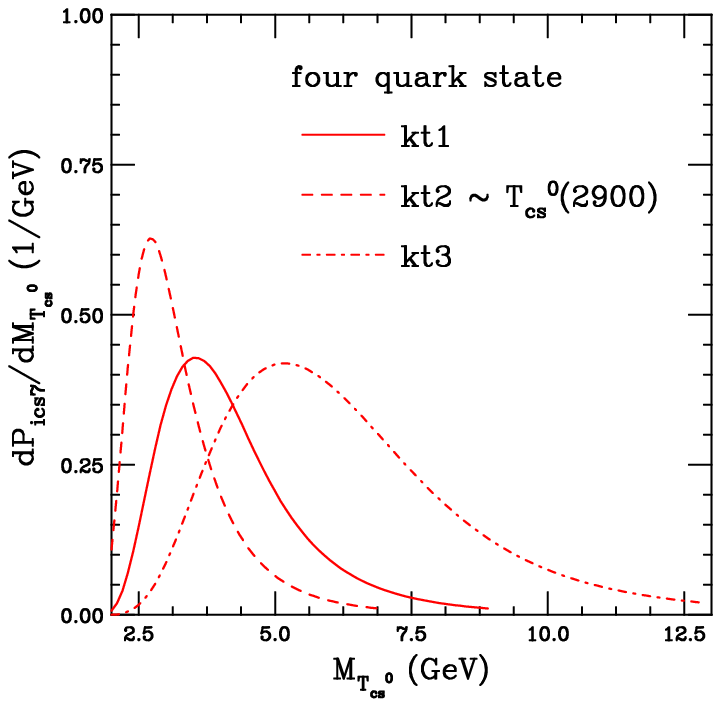}
  \end{center}
  \caption[]{The $T^0_{cs}$ probability distribution, calculated using
    Eq.~(\ref{Tetra_4quark_mass}),
    as a function of mass of the state.
    Calculations are shown for kt1 (solid), kt2 (dashed), and kt3
    (dot-dashed).
  }
\label{ic_Tcs07_Mdists}
\end{figure}

\begin{figure}
  \begin{center}
    \includegraphics[width=0.495\textwidth]{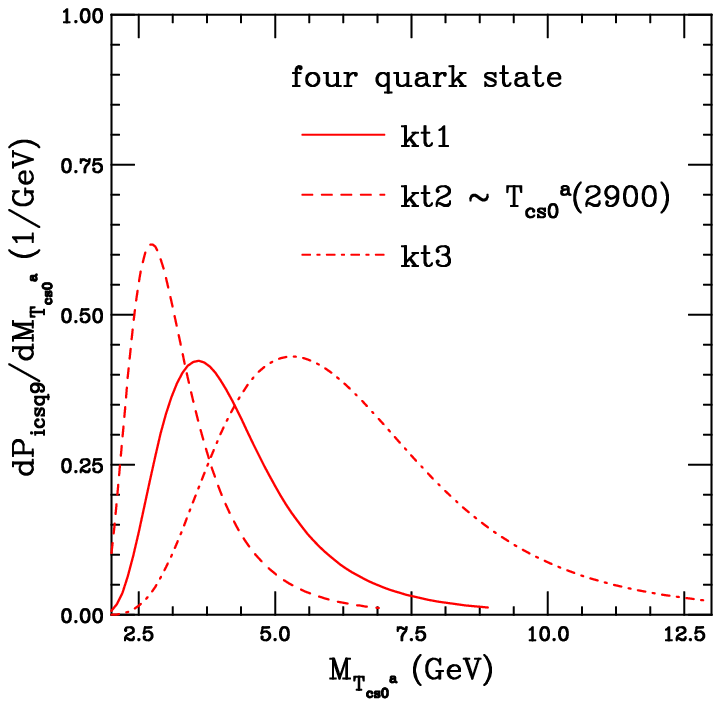}
  \end{center}
  \caption[]{The $T_{cs0}^a$ probability distribution, calculated using
    Eq.~(\ref{Tetra_4quark_mass}),
    as a function of mass of the state.
    Calculations are shown for kt1 (solid), kt2 (dashed), and kt3
    (dot-dashed).
  }
\label{ic_Tcs09_Mdists}
\end{figure}

Figures~\ref{ic_Tcs07_Mdists} and \ref{ic_Tcs09_Mdists} compare the mass
distributions calculated from a 7-particle state for
$T_{cs0}(2900)^0$ and $T_{cs1}(2900)^0$ (both labeled
$T_{cs}^0$ in the figure)
and from a 9-particle state for
$T^a_{c\overline s 0}(2900)^0$ and 
$T^a_{c \overline s0}(2900)^{++}$ (both labeled $T_{cs0}^a$ in the figure).
Recall that in the former case, the antiparticle partner,
$\overline c d \overline s u$, is calculated since producing a
$c \overline d s \overline u$ state in the
intrinsic charm model requires a sub-leading 11-particle state.

Despite the additional light $q \overline q$ pair required to produce the
$T_{cs0}^a$, the mass distributions in Fig.~\ref{ic_Tcs09_Mdists}
are remarkably similar to those of the
$T_{cs}^0$ in Fig.~\ref{ic_Tcs07_Mdists}, giving nearly equal masses and
very similar widths.  In fact, this should not be particularly surprising
because employing the same quark flavors in the tetraquark state should yield a
similar mass distribution, regardless of how many quarks are in the initial
proton wavefunction.  Only the kinematic distributions depend on the number of
particles in the initial proton Fock configuration, as will be shown in
Sec.~\ref{tetra_y_pt}.

In both cases, the average mass with set kt2 is $\approx 3.3$~GeV, comparable
to but slightly larger than the measured mass of the state.  The widths, with
set kt2, are $\approx 0.9$~GeV,
slightly larger than the width calculated for the $T_{c \overline c s}$.

\begin{table}[htbp]
  \begin{center}
    \begin{tabular}{|c|c|c|c|c|} \hline
State & \multicolumn{4}{|c|}{Mass (GeV)} \\ \hline
& kt1 & kt2 & kt3 & kt4 \\ \hline
\multicolumn{5}{|c|}{Meson pair configuration} \\ \hline
$T_{\psi \psi}$ & 6.933 & 6.358 & 7.637 & -  \\ \hline \hline
$X(3872)$ & 4.303 & 4.021 & 5.236 & - \\ \hline
$X_s$ & 4.475 & 4.215 & 5.404 & 4.892 \\ \hline
$T_{cc}^-$ & 4.349 & 4.054 & 5.394 & - \\ \hline \hline
\multicolumn{5}{|c|}{4-quark configuration} \\ \hline
$T_{c \overline c s}$ & 5.404 & 4.301 & 7.716 & - \\ \hline \hline
$T_{cs}^0$ & 4.218 & 3.263 & 6.261 & - \\ \hline
$T_{cs0}^a$ & 4.272 & 3.288 & 6.394 & - \\ \hline
    \end{tabular}
  \end{center}
  \caption[]{The average tetraquark candidate
    mass from intrinsic charm states for the
    $k_T$ integration range given.  Note that only the $X_s$ calculation uses
    set kt4.  Note also that $T_{cs}^0$ refers to both $T_{cs1}^0$ and $T_{cs0}^0$
    while $T_{cs0}^a$ refers to both $T_{c \overline s0}^{a \, 0}$ and
    $T_{c \overline s0}^{a \, ++}$.}
  \label{ave_mass_table}
\end{table}

\begin{table}[htbp]
  \begin{center}
    \begin{tabular}{|c|c|c|c|c|} \hline
State & \multicolumn{4}{|c|}{Width (GeV)} \\ \hline
      & kt1 & kt2 & kt3 & kt4 \\ \hline
\multicolumn{5}{|c|}{Meson pair configuration} \\ \hline
$T_{\psi \psi}$ & 0.797 & 0.748 & 0.905 & -  \\ \hline \hline
$X(3872)$ & 0.456 & 0.422 & 0.739 & - \\ \hline
$X_s$ & 0.431 & 0.432 & 0.702 & 0.534 \\ \hline
$T_{cc}^-$ & 0.515 & 0.510 & 0.732 & - \\ \hline \hline
\multicolumn{5}{|c|}{4-quark configuration} \\ \hline
$T_{c \overline c s}$ & 1.264 & 0.783 & 2.212 & - \\ \hline \hline
$T_{cs}^0$ & 1.251 & 0.883 & 2.057 & - \\ \hline
$T_{cs0}^a$ & 1.270 & 0.895 & 2.087 & - \\ \hline
    \end{tabular}
  \end{center}
  \caption[]{The width (standard deviation) of the tetraquark candidate
    mass distribution
    from intrinsic charm states for the $k_T$ integration range given.  Note
    that only the $X_s$ calculation uses set kt4. Note also that $T_{cs}^0$
    refers to both $T_{cs1}^0$ and $T_{cs0}^0$
    while $T_{cs0}^a$ refers to both $T_{c \overline s0}^{a \, 0}$ and
    $T_{c \overline s0}^{a \, ++}$.}
  \label{FWHM_table}
\end{table}

Overall, the agreement between the calculations and the measured tetraquark
candidate masses is quite good.  The calculations can also distinguish between
assumptions about the nature of the tetraquarks, as either a pair of $D$ or
$D_s$ mesons or as a more loosely bound 4-quark state.  For example, a
tetraquark with the composition $c \overline c d \overline s$ could be arranged
either as $(c \overline s) + (\overline c d) \equiv D_s^+ + D^-$ or
$(c \overline c) + (d \overline s) \equiv J/\psi + K^0$  In the former case,
where the lighter $d$ or $\overline s$ quark is associated with a heavy charm
quark, the meson pair assumption can be used.  In the latter, case, as
discussed earlier in this section, one can obtain a defined mass peak only if
the state is assumed to be composed of four uncorrelated quarks.

\section{Tetraquark Kinematic Distributions}
\label{tetra_y_pt}

The rapidity and $p_T$ distributions of the states discussed in the previous
section are now calculated.  In this section, for ease of calculation,
instead of integrating over the tetraquark mass, the average masses are used
instead.  As in the previous section, all distributions are
shown normalized to unity to facilitate comparison of the shapes of the
distributions.

The denominator of Eq.~(\ref{icdenom}) ensures that the heaviest quarks in the
state carry the largest fraction of the momentum.  This can be manifested by the
charm quarks either carrying a larger fraction of the longitudinal momentum,
represented by Feynman $x$, or $x_F$, or larger transverse momentum.  The number
of quark in the state also plays a role in their kinematic distributions:  when
the available momentum is distributed among more partons, the average phase
space available to each one is reduced.  

\begin{figure}
  \begin{center}
    \includegraphics[width=0.495\textwidth]{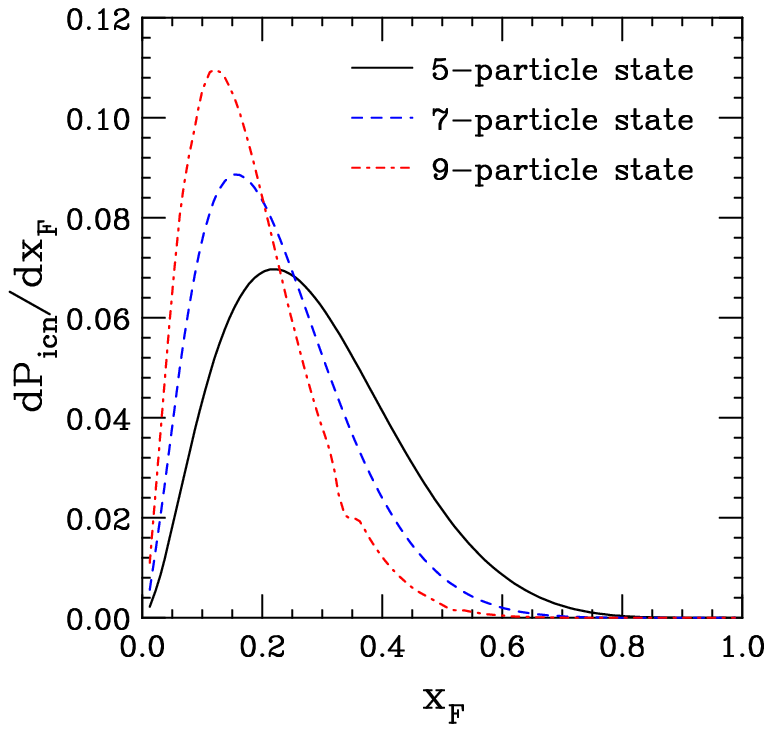}
  \end{center}
  \caption[]{The probability distribution as a function of $x_F$ for
    charm quark production from 5- (solid black), 7- (dashed blue), and
    9-particle states (dot-dashed red).  All calculations use set kt2.
  }
\label{ic_cq_xFdist}
\end{figure}

\begin{figure}
  \begin{center}
    \includegraphics[width=0.495\textwidth]{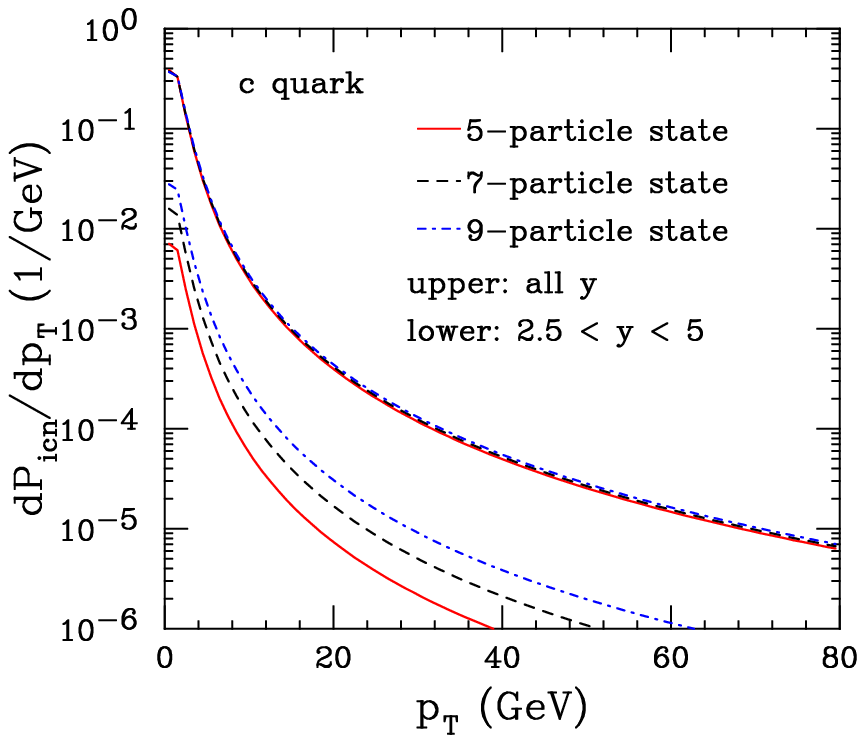}
  \end{center}
  \caption[]{The probability distribution as a function of $p_T$ for
    charm quark production from 5- (solid black), 7- (dashed blue), and
    9-particle states (dot-dashed red).  All calculations use set kt2.
    The upper set of curves are integrated over all rapidity while the
    lower curves correspond to $2.5 < y < 5$ for $\sqrt{s} = 7$~TeV.
  }
\label{ic_cq_pTdist}
\end{figure}

To demonstrate how the number of quarks in the state might affect the kinematic 
distributions, in Figs.~\ref{ic_cq_xFdist} and \ref{ic_cq_pTdist}, the charm
quark Feynman $x$ and $p_T$ distributions are shown for 5-, 7- and 9-particle
states of the proton.
To standardize the result, only light $q \overline q$ pairs are added to the 
$|uud c\overline c \rangle$ state, {\it i.e.}\
$|uud c \overline c q \overline q \rangle$ and
$|uud c \overline c q \overline q q \overline q \rangle$.  Note that $q$ can
represent $u$ or $d$ quarks interchangeably because they are assumed to have the
same current quark mass in the model.

The $x_F$ distribution is shown in Fig.~\ref{ic_cq_xFdist} because it is
independent of
energy and $k_T$ range while the rapidity distribution is not.  Because
$x_F = (2m_T/\sqrt{s})\sinh y$, for fixed $m_T$ and $y$, the $x_F$ value probed
appears at higher rapidity with increasing $\sqrt{s}$.  For example, if
$x_F = 0.2$, near the average $x$ of a charm quark in a 5-particle state,
and $m_T = 2$~GeV, $y = \sinh^{-1}(0.05 \sqrt{s})$ so that for
$\sqrt{s} = 20$~GeV, in the range of prior fixed-target experiments measuring
$J/\psi$ and open charm; 100~GeV, near the top energy for $p+{\rm Ar}$
measurements using the SMOG device at the LHC; and 5~TeV, in the lower range of
LHC $p+p$ collisions; the $y$ value corresponding to this $x_F$ is 0.9, 2.3 and
6.2 respectively.  The effective forward boost of calculating the intrinsic charm distribution as a function of rapidity over a wide range of $\sqrt{s}$ is
illustrated in Ref.~\cite{RV_IC_EN}.  Note also that the rapidity
distributions, as shown later, can also depend on the $k_T$ range of the
integration because $x_F \propto m_T$.

The average $x_F$ of the charm quark decreases as the number of quarks in the
state.  The average $x_F$ of a charm quark in a 5-particle state of 0.285.
If one adds only light $q \overline q$ pairs, the average $x_F$ of a
charm quark decreases to 0.220 for a 7-quark state
and 0.178 for a 9-particle state.  Note that, as a function of rapidity, the
boosted distributions would retain the same hierarchy, with charm quarks from a
5-particle state at higher average rapidity than those from a 7- or 9-particle
state.

The average transverse momentum of a charm quark, integrated over all $x_F$ and
rapidity, slightly increases
with the number of quarks in the state, The increase is likely because the
high $p_T$ tail
of the charm distribution is slightly harder for states with more particles:
$\langle p_T \rangle = 2.32$~GeV for a 5-quark state; 2.37~GeV for a 7-quark
state; and 2.40~GeV for a 9-quark state.  The charm distribution becomes harder
because the additional light quarks in the more populated states have a lower
maximum $k_T$ range.  Note that if no kinematic constraints are included,
such as the finite rapidity acceptance of a detector, the $p_T$ distribution
would be independent of energy \cite{RV_IC_EN}.  Thus, at low
fixed-target energies,
the $p_T$ distribution from intrinsic charm is broader that that from
perturbative QCD but at higher energies, $\sqrt{s} > 40$~GeV, the two
contributions become difficult to distinguish at higher $p_T$ when no rapidity
cuts are considered \cite{RV_IC_EN}. So far, poor experimental statistics
at high $p_T$ limit the potential determination of a large intrinsic charm
contribution in this part of phase space, see the discussion in
Ref.~\cite{RV_SMOG} for example.

As further shown in Ref.~\cite{RV_IC_EN},
the $J/\psi$ and $\overline D$ meson $p_T$ distributions from intrinsic
charm depend on the calculated rapidity range.  This is demonstrated in
Fig.~\ref{ic_cq_pTdist} where the charm quark
$p_T$ distribution integrated over all
$x_F$ ($0 < x_F < 1$) is compared to that for an $x_F$ range appropriate for
$2.5 < y < 5$ at $\sqrt{s} = 7$~TeV, corresponding to $0.00454 < x_F < 0.0557$.
As can be seen
in Fig.~\ref{ic_cq_xFdist}, because the average $x_F$ is reduced for states
with more quarks, the $p_T$ distribution from a 9-particle state encompasses
more of the $p_T$ distribution at this low $x_F$ than the 5- or 7-particle
states.  The energy dependence of the $p_T$ distribution is discussed in
more detail in Sec.~\ref{Tetra_pTdists}.

The assumption of a fixed tetraquark mass in the calculation of the $y$ and
$p_T$ distributions, in contrast to the mass distributions, means that the
calculations are independent of whether the tetraquark
is composed of a meson pair or four uncorrelated quarks.  Thus the
momentum-conserving delta functions and the additional delta functions needed
to connect the four constituents of the tetraquark are all that are required in
addition to Eq.~(\ref{icdenom}). Thus the
kinematic distributions are independent of any correlations between the
partons in the state as long as the same number and type of partons are
included.  This observation was also made in Ref.~\cite{ANDY}.

With this starting point, the rapidity and $p_T$ distributions are shown in Secs.~\ref{Tetra_ydists} and \ref{Tetra_pTdists} respectively.

\subsection{Rapidity Distributions}
\label{Tetra_ydists}

\begin{table}[htbp]
  \begin{center}
    \begin{tabular}{|c|c|c|c|} \hline
State & \multicolumn{3}{|c|}{$\sqrt{s}$} \\ \hline
      & 5~TeV & 7~TeV & 13~TeV \\ \hline
$T_{\psi \psi}(6600)$ & 5.96 & 6.30 & 6.91 \\ \hline
$T_{\psi \psi}(6900)$ & 5.40 & 6.24 & 6.85 \\ \hline \hline 
$X(3872)$ & 6.46 & 6.80 & 7.42 \\ \hline
$X_s$(kt1) & 6.32 & 6.66 & 7.27 \\ \hline
$X_s$(kt2) & 6.42 & 6.76 & 7.39 \\ \hline
$X_s$(kt3) & 6.06 & 6.39 & 7.02 \\ \hline
$X_s$(kt4) & 6.21 & 6.54 & 7.15 \\ \hline
$T_{cc}^-$ & 6.27 & 6.61 & 7.23 \\ \hline
$T_{c \overline c s}$ & 6.42 & 6.77 & 7.38 \\ \hline \hline 
$T_{cs}^0$ & 6.56 & 6.40 & 7.51 \\ \hline
$T_{cs0}^a$ & 6.34 & 6.68 & 7.30 \\ \hline
    \end{tabular}
  \end{center}
  \caption[]{The average tetraquark candidate
    rapidity from intrinsic charm states
    for different LHC energies.  Note that $T_{cs}^0$
    refers to both $T_{cs1}^0$ and $T_{cs0}^0$
    while $T_{cs0}^a$ refers to both $T_{c \overline s0}^{a \, 0}$ and
    $T_{c \overline s0}^{a \, ++}$.}
  \label{ave_rapidity_table}
\end{table}

The rapidity distributions are calculated for $\sqrt{s} = 5$, 7 and 13~TeV,
all center of mass energies for $p+p$ collisions at the LHC.  To facilitate comparison between tetraquark states, typically only results are shown for
$\sqrt{s} = 7$~TeV.  The average value of the rapidity is given for all
energies in Table~\ref{ave_rapidity_table}.

\begin{figure}
  \begin{center}
    \includegraphics[width=0.495\textwidth]{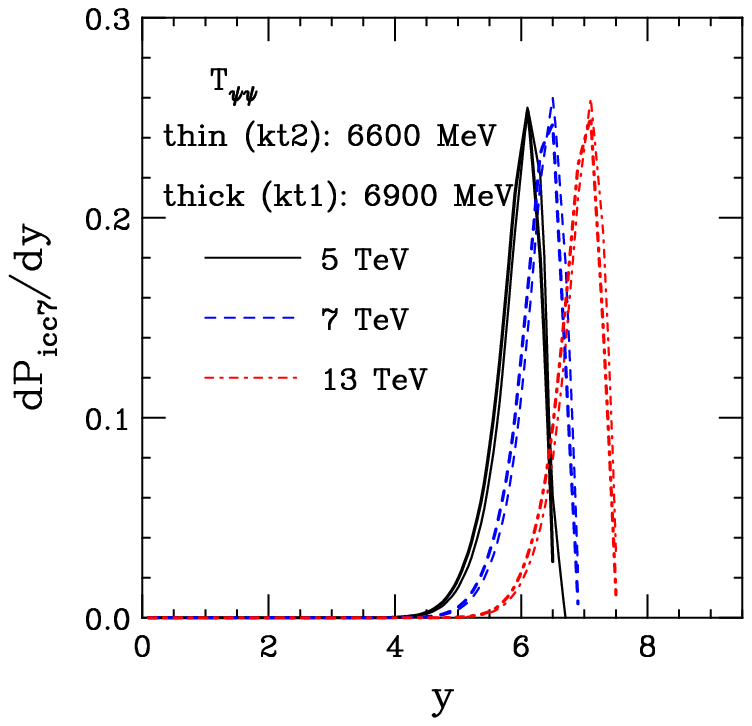}
  \end{center}
  \caption[]{The probability distribution as a function of rapidity for
    $T_{\psi \psi}$ production
    at $\sqrt{s} = 5$ (solid black), 7 (dashed blue), and 13~TeV (dot-dashed
    red).  The thin lines, for $T_{\psi \psi}(6600)$, use set kt2 while the thick
    lines, for $T_{\psi \psi}(6900)$, use set kt1.
  }
\label{ic_Tpsipsi_ydists}
\end{figure}

Figure~\ref{ic_Tpsipsi_ydists} shows the  rapidity distributions for the
$T_{\psi\psi}(6600)$ and $T_{\psi\psi}(6900)$, obtained with the $k_T$ ranges with
sets kt2 and kt1 respectively.  As expected, the distributions peak at
rapidities greater than 5 with the furthest forward distributions being those at
$\sqrt{s} = 13$~TeV.  Because the $T_{\psi \psi}$ has the largest mass of all the
tetraquark candidate states considered, it is boosted least. 

The differences between the distributions based on mass ($k_T$ range) are
negligible on the scale of the plots.  They are generally less than 0.1 unit of
rapidity given the 300~MeV difference in mass.  

All the distributions in this figure and, indeed, all the rapidity distributions
shown in this section, have a characteristic shape due to the transformation
from $x_F$ to rapidity.  There is a long tail from zero rapidity until just
below the peak where the distribution abruptly climbs.  The descent to the edge
of phase space above the peak is very abrupt, nearly vertical. This seems almost
counterintuitive compared to the charm quark $x_F$ distribution where the rise
from $x_F =0$ to the peak is faster than that descent above it as
$x_F \rightarrow 1$.  However, as noted in the discussion of
Fig.~\ref{ic_cq_xFdist}, at $\sqrt{s} = 7$~TeV, while the boost affects all
energies similarly, it is less pronounced at lower rapidity.

\begin{figure}
  \begin{center}
    \includegraphics[width=0.495\textwidth]{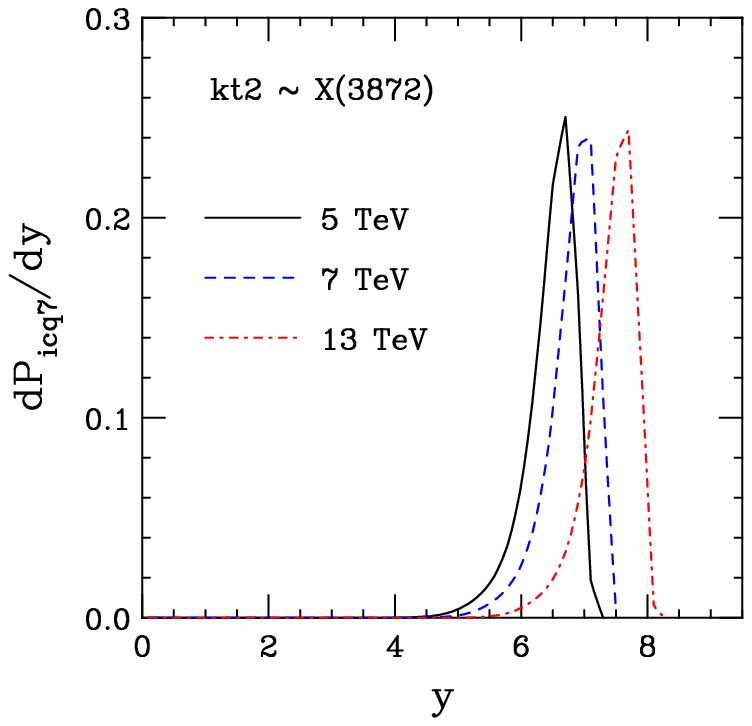}
  \end{center}
  \caption[]{The probability distribution as a function of rapidity for
    $X(3872)$
    production
    at $\sqrt{s} = 5$ (solid black), 7 (dashed blue), and 13~TeV (dot-dashed
    red), all calculated using parameter set kt2.
  }
\label{ic_X_ydists}
\end{figure}

The $X(3872)$ rapidity distributions are shown for all three LHC energies in
Fig.~\ref{ic_X_ydists}.  They are calculated with set kt2, the set of $k_T$
integration ranges that agree best with the measured $X(3872)$ mass.  While the
distributions are similar to those shown in Fig.~\ref{ic_Tpsipsi_ydists}, the
lighter mass of the $X(3872)$ results in a more forward-peaked rapidity
distribution.

\begin{figure}
  \begin{center}
    \includegraphics[width=0.495\textwidth]{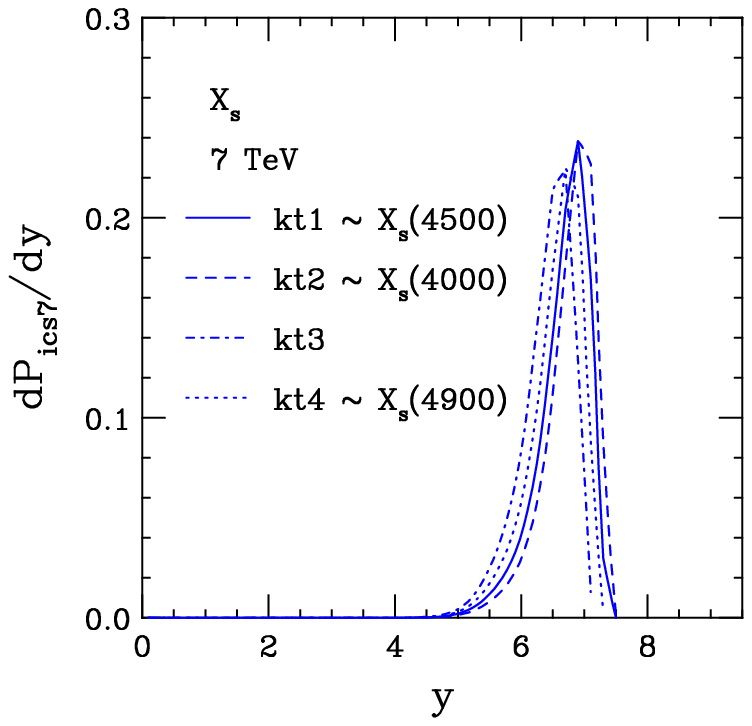}
  \end{center}
  \caption[]{The probability distribution as a function of rapidity for $X_s$
    production
    at $\sqrt{s} = 7$~TeV for parameter sets kt1 (solid), kt2 (dashed), kt3
    (dot dashed) and kt4 (dotted).  The approximately grouped $X_s$
    masses based on
    Table~\ref{tetraquark_table} are associated with the closest $k_T$ range.
  }
\label{ic_Xs_ydists}
\end{figure}

Figure~\ref{ic_Xs_ydists} shows the rapidity distributions for all four sets of
$k_T$ ranges for the $X_s$.  Sets kt2, kt1 and kt4 correspond to the states
measured at approximately 4.0, 4.5 and 4.9~GeV.  The different masses and
the associated $k_T$ ranges affect the rapidity distributions.  The most
forward distribution is found for the lowest mass and narrowest $k_T$ range,
set kt2.  The increased $k_T$ range corresponding to higher masses shifts
the distribution backward in rapidity so that set
kt2 (mass 4.0 GeV) is forward of set kt1 (mass 4.5 GeV), with sets kt4 (mass
4.9 GeV) and kt3 (mass 5.4 GeV, see Table~\ref{ave_mass_table})
peaking at lower rapidities.
The rapidity shift due to the changes in mass and $k_T$ range is on
the order of 0.1
units of rapidity, similar to the difference noted for the $T_{\psi \psi}$.

\begin{figure}
  \begin{center}
    \includegraphics[width=0.495\textwidth]{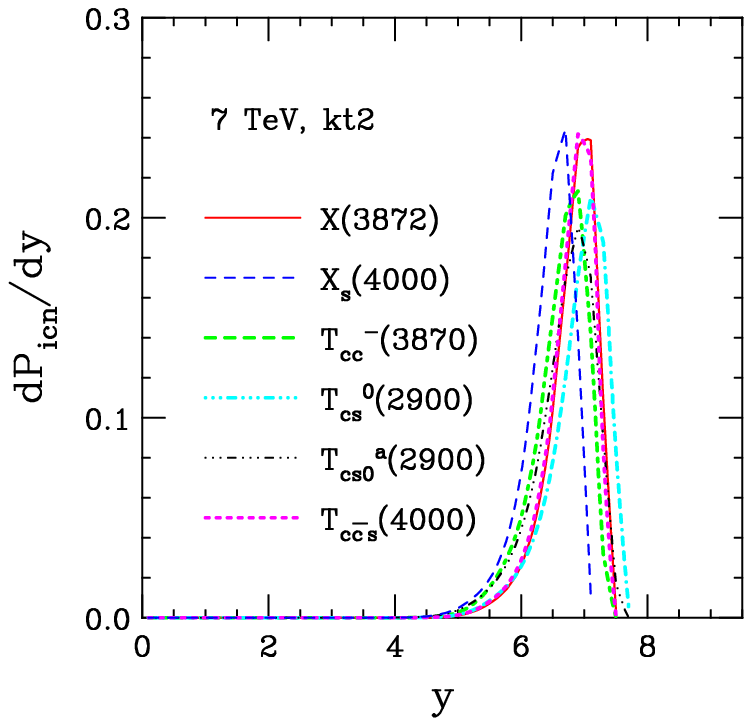}
  \end{center}
  \caption[]{The probability distribution as a function of rapidity at
    $\sqrt{s} = 7$~TeV using parameter set kt2 for $X(3872)$ (solid red),
    $X_s$ (dashed blue), $T_{cc}^-$ (dot-dashed green); $T_{cs0}^a$
    (dot-dot-dot-dashed black), $T_{cs}^0$ (dash-dash-dash-dotted cyan) and
    $T_{c \overline c s}$ (dotted magenta).
  }
\label{ic_Picc7_ydists}
\end{figure}

Figure~\ref{ic_Picc7_ydists} compares all the rapidity distributions for
tetraquark candidates with one or two charm quarks at $\sqrt{s} = 7$~TeV.  All
distributions are calculated for set kt2 so that the $X_s$ distribution shown
corresponds to the lowest mass state.  The $X(3872)$, $T_{cc}^-$, and $T_{c \overline c s}$
all have similar quark content (at least two charm quarks) and masses, thus
their rapidity distributions are also similar.  Note that the exchange of a
light quark for a strange quark in the $T_{c \overline c s}$ does not significantly affect the mass or rapidity.
The $X_s$, with two strange quarks, has the lowest average rapidity. 

The $T_{cs}^0$, with a single charm quark, arising from a 7-particle state, is
the most forward.  It is worth noting that while the quark content is very
similar to that of the $T_{cs0}^a$ -- one charm quark, one strange quark and two
light quarks -- the peak of the $T_{cs0}^a$ is shifted backward by about 0.2
units of rapidity relative to the $T_{cs}^0$.  That is because the $T_{cs0}^a$
must be produced from a 9-particle state.  If a`non-leading' 11-particle states
had been considered, e.g. to produce the $T_{cc}^+$ instead of the $T_{cc}^-$,
one could expect a similar small
backward shift for rapidity distributions from such
states even though the quark content and mass are similar.  This is clear from
Fig.~\ref{ic_cq_xFdist}.   

\subsection{Transverse Momentum Distributions}
\label{Tetra_pTdists}

The $p_T$ distributions are now shown for the tetraquark candidates, under
the same conditions as in Sec.~\ref{Tetra_ydists}.

\begin{table}[htbp]
  \begin{center}
    \begin{tabular}{|c|c||c|c||c|c||c|c|} \hline
      & & \multicolumn{2}{|c|}{$\sqrt{s} = 5$~TeV} &
      \multicolumn{2}{|c|}{$\sqrt{s} = 7$~TeV} &
      \multicolumn{2}{|c|}{$\sqrt{s} = 13$~TeV} \\ \hline
      & all $y$ & \multicolumn{6}{|c|}{$2.5 < y < 5$} \\ \hline
      State & $\langle p_T \rangle$
      & \%$P_{{\rm ic}, n}$ & $\langle p_T \rangle$ 
      & \%$P_{{\rm ic}, n}$ & $\langle p_T \rangle$ 
      & \%$P_{{\rm ic}, n}$ & $\langle p_T \rangle$ \\ \hline
$T_{\psi \psi}(6600)$ & 4.05 & 3.40 & 21.74 & 1.52 & 30.32 & 0.33 & 47.50 \\ \hline
$T_{\psi \psi}(6900)$ & 4.56 & 4.52 & 21.92 & 2.04 & 30.45 & 0.45 & 47.61 \\ \hline \hline
$X(3872)$ & 2.82 & 1.36 & 23.94 & 0.65 & 31.82 & 0.15 & 48.20 \\ \hline
$X_s$(kt1) & 3.49 & 2.18 & 24.19 & 1.03 & 32.02 & 0.24 & 48.24 \\ \hline
$X_s$(kt2) & 2.97 & 1.53 & 24.01 & 0.73 & 31.86 & 0.17 & 48.13 \\ \hline
$X_s$(kt3) & 4.75 & 4.36 & 24.18 & 2.07 & 32.12 & 0.48 & 48.42 \\ \hline
$X_s$(kt4) & 4.09 & 3.10 & 24.33 & 1.50 & 32.16 & 0.35 & 48.37 \\ \hline
$T_{cc}^-$ & 3.93 & 2.76 & 24.18 & 1.34 & 31.96 & 0.31 & 48.15 \\ \hline
$T_{c \overline c s}$ & 2.98 & 1.92 & 24.81 & 0.93 & 32.11 & 0.21 & 48.29 \\ \hline \hline
$T_{cs}^0$ & 3.22 & 1.83 & 24.91 & 0.90 & 32.46 & 0.21 & 48.39 \\ \hline
$T_{cs0}^a$ & 3.19 & 2.94 & 19.92 & 1.47 & 26.48 & 0.38 & 41.22 \\ \hline
    \end{tabular}
  \end{center}
  \caption[]{The average tetraquark candidate
    $p_T$ (in GeV) from intrinsic charm states
    for $p+p$ collisions at $\sqrt{s} = 5$, 7 and 13 TeV.  At each energy, the
    percentage of the total $p_T$ distribution captured in the rapidity range
    $2.5 < y < 5$ is given
    along with the average $p_T$ at that energy.  The average $p_T$ integrated
    over all rapidity, independent of energy, is also shown.
    Note also that $T_{cs}^0$ refers to both $T_{cs1}^0$ and $T_{cs0}^0$
    while $T_{cs0}^a$ refers to both $T_{c \overline s0}^{a \, 0}$ and
    $T_{c \overline s0}^{a \, ++}$.}
  \label{ave_pT_table}
\end{table}

The $p_T$ distributions are again calculated for $\sqrt{s} = 5$, 7 and 13~TeV,
all center of mass energies for $p+p$ collisions at the LHC.  To facilitate
comparison between tetraquark states, typically only results are shown for
$\sqrt{s} = 7$~TeV.  The average $p_T$ value is given for all energies in
Table~\ref{ave_pT_table}.  The averages are first given for the entire forward
rapidity range and then assuming that the rapidity range covered is
$2.5 < y < 5$.  Because the amount of the total $p_T$ distribution captured
depends on the rapidity range, the percentage of the total probability for that
energy and rapidity range is given as \%$P_{{\rm ic}, n}$.

While the average $p_T$ calculated over all rapidities is rather moderate and
similar to the average $p_T$ of the charmonium states, albeit somewhat higher,
when a finite rapidity region is considered, the average $p_T$ increases by an
order of magnitude and grows with center of mass energy while the percentage of
the $p_T$ distribution captured by the rapidity range decreases as the
tetraquark candidate is boosted further forward in rapidity as $\sqrt{s}$
increases.

\begin{figure}
  \begin{center}
    \includegraphics[width=0.495\textwidth]{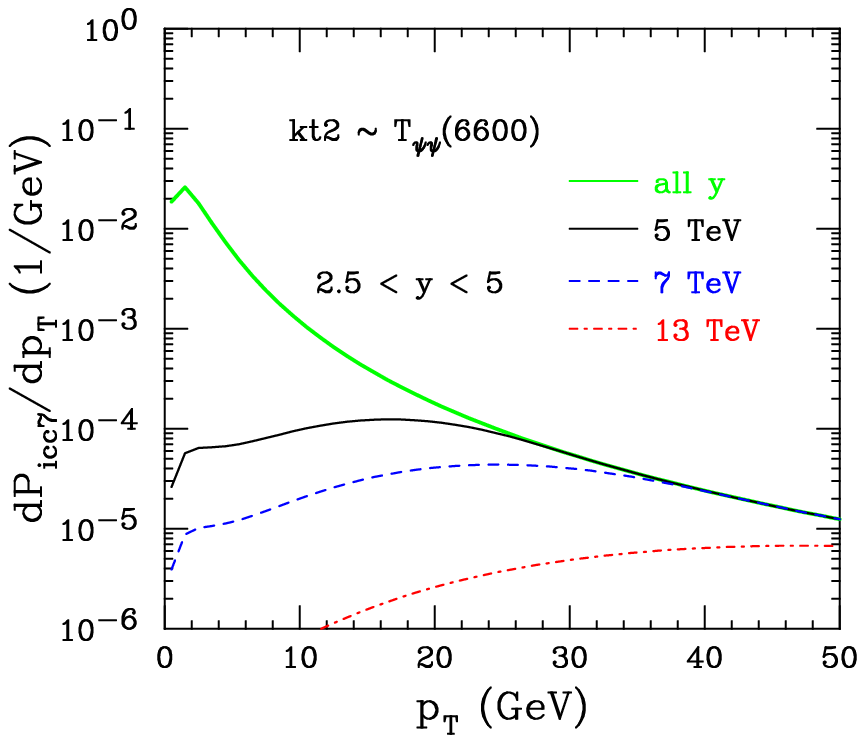}
  \end{center}
  \caption[]{The probability distribution as a function of $p_T$ for
    $T_{\psi\psi}(6600)$
    production
    at $\sqrt{s} = 5$ (solid black), 7 (dashed blue), and 13~TeV (dot-dashed
    red), all calculated using parameter set kt2, corresponding to the mass of 
    $\sim 6600$~MeV.
  }
\label{ic_Tpsipsi_pTdists}
\end{figure}

\begin{figure}
  \begin{center}
    \includegraphics[width=0.495\textwidth]{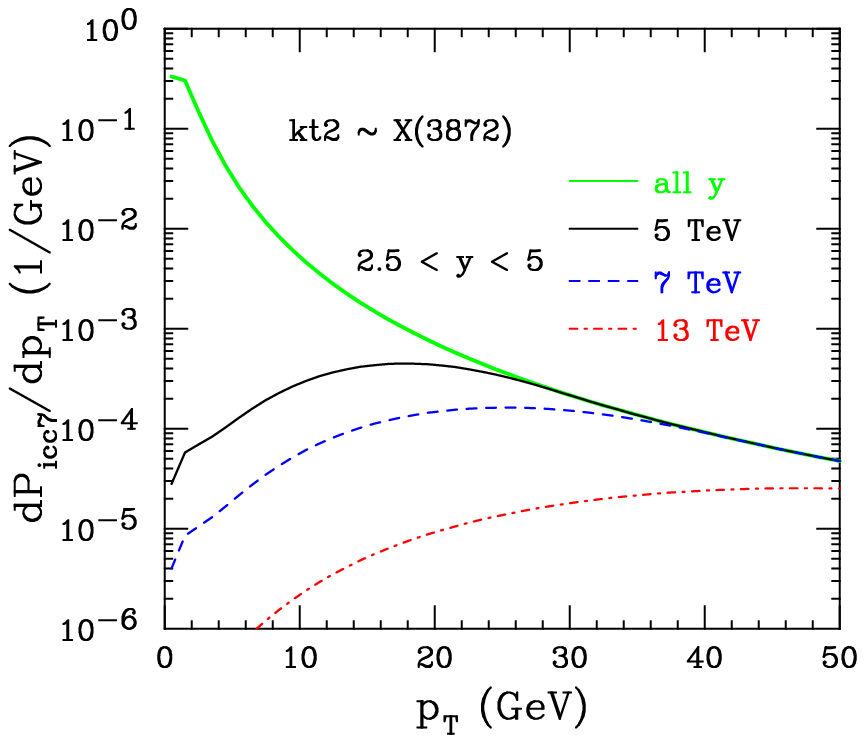}
  \end{center}
  \caption[]{The probability distribution as a function of $p_T$ for $X(3872)$
    production
    at $\sqrt{s} = 5$ (solid black), 7 (dashed blue), and 13~TeV (dot-dashed
    red), all calculated using parameter set kt2.
  }
\label{ic_X_pTdists}
\end{figure}

The reason for this is illustrated in Figs.~\ref{ic_Tpsipsi_pTdists} and
\ref{ic_X_pTdists}.  As shown in Ref.~\cite{RV_IC_EN}, the relation between the
Feynman $x$ of the hadron created by coalescence of the constituent quarks in
the $J/\psi$, $\overline D$ meson, or charm tetraquark candidate, and rapidity
means that, for a fixed value of $x_F$, the maximum $p_T$ can be quite large
according to the definition $x_F = (2m_T/\sqrt{s})\sinh y$.  

Because the tetraquark candidate in the intrinsic charm picture is comoving with
the parent proton, it can manifest itself at rather high $p_T$, even at
relatively high rapidity.  For example, if $x_F= 1$ and $m_T = 1$~GeV, the
maximum rapidity is $\sinh^{-1}(\sqrt{s}/2) = 8.85$ with $\sqrt{s} = 7$~TeV.  If
$m_T = \sqrt{s}/2$, then the maximum rapidity is $y =\sinh^{-1}(1) = 0.881$,
near midrapidity, for the same energy.   At $x_F \sim 0$, on the other hand,
then $m_T \sim 0$ for any rapidity.  Thus in Figs.~\ref{ic_Tpsipsi_pTdists} and
\ref{ic_X_pTdists}, the $p_T$ distribution, integrated over all rapidity, peaks
at $p_T \sim 0$ and then decreases slowly with $p_T$ until near the edge of
phase space.  

If, on the other hand, one considers a finite rapidity range, the distributions
can behave quite differently, as shown in these figures at forward rapidity for
$\sqrt{s} = 5$, 7 and 13~TeV.  The maximum $m_T$ in the forward rapidity range
covered by LHCb, $2.5 < y < 5$, is 578.5~GeV for $y=2.5$ and 47.2~GeV for $y=5$,
assuming $x_F \equiv 1$.  Thus the low $p_T$ part of the $p_T$ spectrum is
suppressed at forward rapidity.  Increasing the center of mass energy from
$\sqrt{s}$ from 5 to 13~TeV consequently increases the suppression of the low
$p_T$ spectral contribution.  However, as the $p_T$ increases, the suppression
is reduced until, at sufficiently high $p_T$ (higher for larger $\sqrt{s}$), the
spectrum is no longer suppressed and the distributions merge with that of the
rapidity-integrated spectrum.  This low $p_T$ spectral suppression in a finite
rapidity range leads to the large increase in the average $p_T$ seen in
Table~\ref{ave_pT_table}.  Assuming a lower rapidity range at the same center of
mass energies would lead to greater suppression at low $p_T$ and increase the
average $p_T$ still further.  The maximum $m_T$ is reduced for lower values of
$x_F$.  

Because $m_T = \sqrt{p_T^2 + m^2}$, for fixed $m_T$, a larger mass particle
reduces the $p_T$ range.  The difference in the maximum $p_T$ between the
$T_{\psi \psi}$ mass of 6.6 or 6.9~GeV and that of the $X(3872)$ is small.
However,one can observe a change in the spectral shapes for the $T_{\psi \psi}$
and the $X(3872)$ and, indeed, greater low $p_T$ suppression for the more
massive state.

In the remainder of this section, only distributions at $\sqrt{s} = 7$~TeV and
in the rapidity range $2.5 < y < 5$ are shown to illustrate differences in the
chosen $k_T$ ranges (for $X_s$ production)
and the general makeup of the states themselves.  

\begin{figure}
  \begin{center}
    \includegraphics[width=0.495\textwidth]{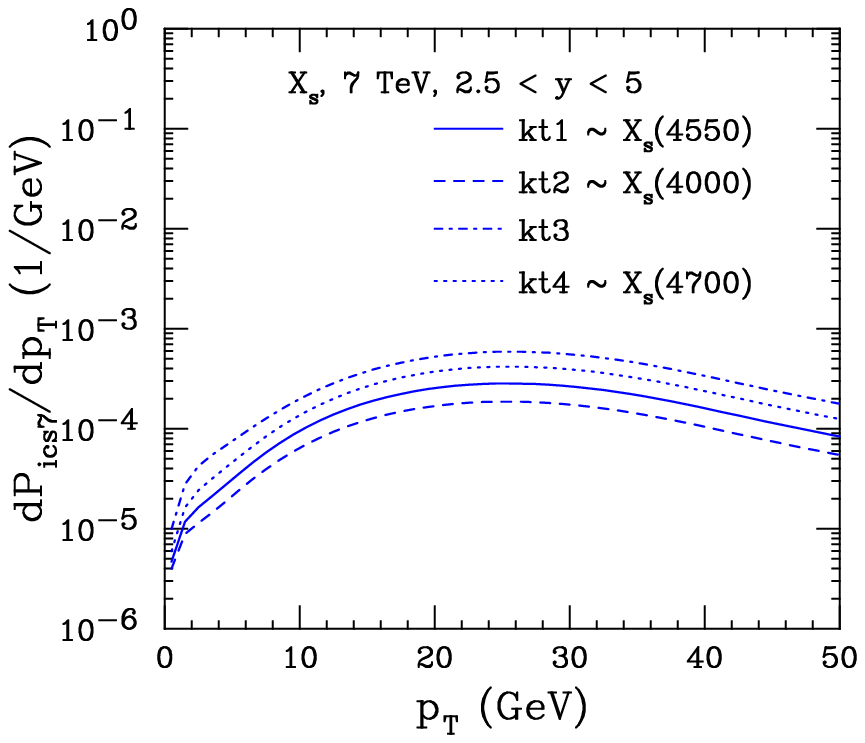}
  \end{center}
  \caption[]{The probability distribution as a function of rapidity for $X_s$
    production
    at $\sqrt{s} = 7$~TeV for parameter sets kt1 (solid), kt2 (dashed), kt3
    (dot dashed) and kt4 (dotted).  The approximately grouped $X_s$
    masses based on
    Table~\ref{tetraquark_table} are associated with the closest $k_T$ range.
  }
\label{ic_Xs_pTdists}
\end{figure}

Figure~\ref{ic_Xs_pTdists} shows the  $p_T$ distributions at 7~TeV for the $X_s$
states, including all four sets of $k_T$ ranges, each calculated with the
average mass given in Table~\ref{ave_mass_table}.  It is clear that the
larger $k_T$ range and corresponding larger mass
results in a somewhat harder $p_T$ distribution with a slightly
higher peak.  In the case of the lowest mass, kt2,
the percentage of the $p_T$ distribution captured in the rapidity range
$2.5 < y < 5$, given as \%$P_{{\rm ic}, n}$, is the smallest with increasing
fractions for sets kt1, kt4 and kt3 respectively, corresponding to broader
$k_T$ integration ranges.  The average $p_T$ integrated over all rapidity also
increases with increasing $k_T$ range and mass of the state,
as observed in Table~\ref{ave_pT_table}.
The change in the average $p_T$ is much smaller when the rapidity range is
fixed to $2.5 < y < 5$.  

\begin{figure}
  \begin{center}
    \includegraphics[width=0.495\textwidth]{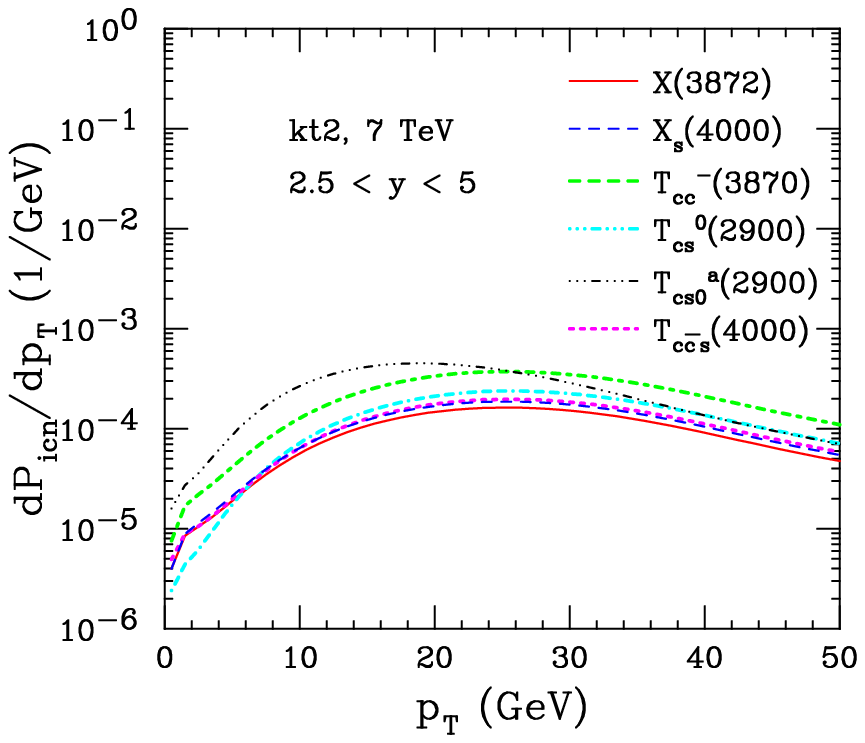}
  \end{center}
  \caption[]{The probability distribution as a function of $p_T$ in the rapidity
    interval $2.5 < y < 5$ at
    $\sqrt{s} = 7$~TeV using parameter set kt2 for $X(3872)$ (solid red),
    $X_s$ (dashed blue), $T_{cc}^-$ (dot-dashed green); $T_{cs1}$ (dot-dot-dot-dashed
    black), $T_{cs0}$ (dash-dash-dash-dotted cyan) and $T_{c \overline c s}$
    (dotted magenta).
  }
\label{ic_Picc7_pTdists}
\end{figure}

Finally, the $p_T$ distributions for tetraquark candidates with one and two
charm quarks are compared in Fig.~\ref{ic_Picc7_pTdists}, also at
$\sqrt{s} = 7$~TeV.  Here all the distributions are calculated with set kt2.
Integrated over rapidity, the average $p_T$ values for
states with two charm quarks are all around 3~GeV.  The average $p_T$ of the
$T_{cs}^0$ and $T_{cs0}^a$, with a single charm quark, are somewhat larger but
the difference is small.  When the rapidity cut is applied,  the percentage of
the distributions captured are all similar for the $X(3872)$, $X_s$, $T_{cc}^-$,
$T_{c \overline c s}$ and $T_{cs}^0$.  However, the values of \%$P_{{\rm ic}, n}$ are larger
overall for the $T_{cs0}^a$ because it is produced from a 9-particle state while
the others are all produced from 7-particle states.  The average $p_T$ of the
distribution within the rapidity range is also higher, see
Table~\ref{ave_pT_table}.  The backward shift to lower $p_T$ of the $T_{cs0}^a$
distribution is also clear in Fig.~\ref{ic_Picc7_pTdists}.

\section{Estimated Cross Sections}
\label{Rates}

In this section, the estimated cross sections are briefly discussed.
The production
cross section for a single intrinsic $c \overline c$ pair from a 5-particle
$|uudc \overline c \rangle$ configuration of the proton can be written as 
\cite{Vogt:1994zf}
\be
\sigma_{{\rm ic}\, 5}(pp) = P_{{\rm ic}\, 5}^0 \sigma_{p N}^{\rm in}
\frac{\mu^2}{4 \widehat{m}_c^2} \, \, .
\label{icsign}
\ee
The resolving factor of $\mu^2/4 \widehat{m}_c^2$ arises from a soft
interaction, breaking the coherence of the Fock state \cite{BHMT}.
This factor was introduced to calculate the cross section for intrinsic charm
in Ref.~\cite{Vogt:1994zf}.
Here 
$\mu^2 = 0.1$~GeV$^2$ is assumed \cite{RV_SeaQuest}.
The inelastic
cross section $\sigma_{pN}^{\rm in} = 30$~mb is appropriate for fixed-target
interactions and can be used as a conservative value here.  Thus, for
$P_{{\rm ic}, 5}^0$ between 0.1\% and 1\%,
$0.231 \leq \sigma_{{\rm ic}\, 5}(pp) \leq 2.31$~$\mu$b. This range is compatible
with the cross section predicted in Ref.~\cite{Vogt:1994zf}.

The cross section for double $J/\psi$ production was estimated in
Ref.~\cite{dblic} based on the NA3 double $J/\psi$ production measurements in
$\pi^- + N$ interactions at pion beam energies of 150 and 280~GeV \cite{Badpi}
and $p+N$ interactions with a proton beam of 400~GeV \cite{Badp}.  Based on the
cross sections obtained from these data, for the lowest double intrinsic
$c \overline c$ state, a 6-particle
$|\overline u d c \overline c c \overline c \rangle$ for a $\pi^-$ beam and a
7-particle $|u u d c \overline c c \overline c \rangle$ for a $p$ beam,
$P_{{\rm icc}, 7}^0$ is between 4.4\% and 20\% of $P_{{\rm ic}, 5}^0$.  Then, for
$T_{\psi \psi}$, $X(3872)$, and $T_{cc}^-$, all based on a 7-particle state with
two intrinsic $c \overline c$ pairs, the production cross section is
\be
\sigma_{{\rm icc}\, 7}(pp) = P_{{\rm icc}\, 7}^0 \sigma_{p N}^{\rm in}
\frac{\mu^2}{4 \widehat{m}_c^2} \, \, .
\label{ic7sign}
\ee
Note that the same soft interaction factor is assumed for the 7-particle state
and the 5-particle state.  Then, given the same range of $P_{{\rm ic}, 5}^0$ as
above, $10.2 \leq \sigma_{{\rm icc}\, 7}(pp) \leq 465$~nb, based on the lower and
upper bounds of $P_{{\rm icc}, 7}^0$ relative to $P_{{\rm ic}, 5}^0$. 

On the other hand, for $X_s$, $T_{c \overline c s}$, and $T_{cs}^0$, the second intrinsic
$c \overline c$ pair is replaced by an $s \overline s$ pair.  One could
conservatively assume that the probability $P_{{\rm ics}, 7}$ is larger than
$P_{{\rm icc}, 7}$ by the ratio $(\hat m_c^2/\hat m_s^2) \sim 10$, giving
$0.10 \leq \sigma_{{\rm ics}\, 7}(pp) \leq 4.6$~$\mu$b.

Finally, the $T_{cs0}^a$ ($T_{c\overline s0}^0$ and $T_{c \overline s0}^{++}$) is
based on 9-particle Fock states
with intrinsic $c \overline c$, $s \overline s$ and $q \overline q$ pairs where
$q$ represents a light quark.  Given that the additional $q \overline q$ pair
is light, there could be no additional penalty so that
$P_{{\rm icsq}, 9}^0 \approx P_{{\rm ics}, 7}^0$ is a not unreasonable first
approximation, leading to
$\sigma_{{\rm icsq}\, 9}(pp) \approx \sigma_{{\rm ics}\, 7}(pp)$.

These cross sections should be taken as estimates of upper limits and not as
exact predictions.  First, these cross sections encompass all possible charm
hadron combinations that could be derived from these 5-, 7- and 9-particle
configurations of the proton.  These combinatorics have not been worked out
here but would reduce the cross sections of individual tetraquarks.  Next, if
one or more of the components of the tetraquark emerges as a $J/\psi$, the cross
section has to be reduced by a factor of $F_C \sim 2$\% for each emergent
$J/\psi$.  Thus the $T_{\psi \psi}$ cross section could be reduced by
$F_C^2 \sim 4 \times 10^{-4}$ to between 4.1 and 190~pb while a state like the
$T_{c \overline c s}$ would have the cross section reduced by a single factor of $F_C$.
These total cross sections do not include any reduction due to phase space
acceptance, which can be significant, see Table~\ref{ave_pT_table}.  Finally, it
is unclear how much, if at all, the configuration or size of the tetraquark
state might affect its cross section, in addition to its width.

\begin{figure}
  \begin{center}
    \includegraphics[width=0.495\textwidth]{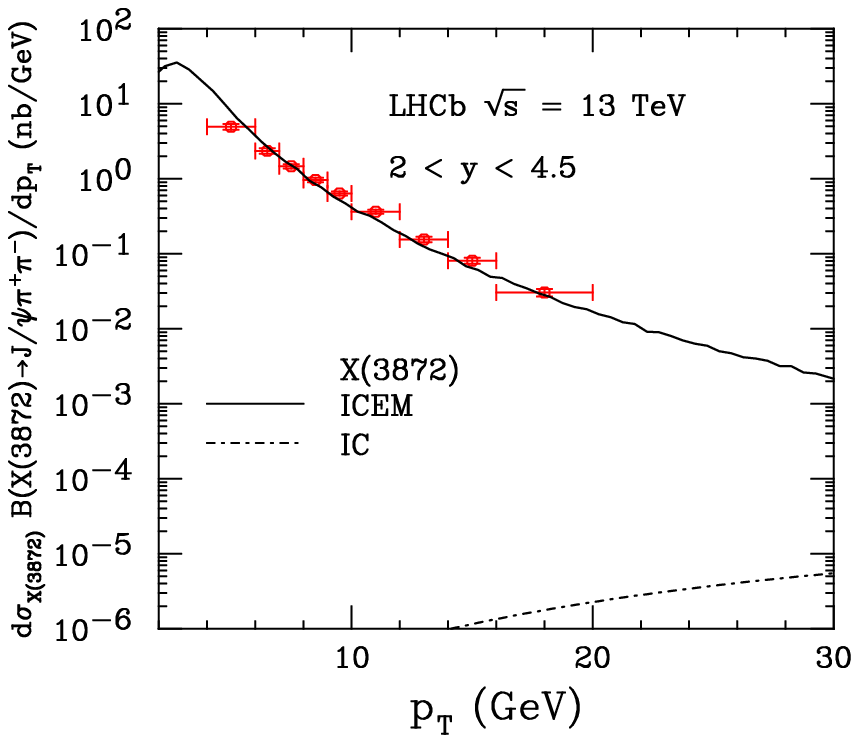}
  \end{center}
  \caption[]{The cross section of $X(3872)$ production as a function of
    $p_T$ as measured by LHCb \cite{LHCb_XofpT} (red points) at
    $\sqrt{s} = 13$~TeV and $2<y< 4.5$.
    The distribution calculated in the ICEM is shown by the solid line while
    the dot-dashed line gives the contribution from intrinsic charm in the
    rapidity range at this energy.
  }
\label{X3872_LHCb}
\end{figure}

As an example of the potential contribution of intrinsic charm to
tetraquark
production, Fig.~\ref{X3872_LHCb} shows the $X(3872)$ $p_T$ distribution at
$\sqrt{s} = 13$~TeV in the rapidity range $2 < y < 4.5$, somewhat shifted to
lower rapidity
relative to the calculations shown previously.  The data from LHCb
\cite{LHCb_XofpT} are shown by the red points.  The solid curve shows an
improved color evaporation model (ICEM) prediction 
prediction with the mass range adjusted to encompass the $X(3872)$ mass.  Since
the $X(3872)$ mass is only $\sim 200$~MeV larger than that of the $\psi$(2S),
the $p_T$ distributions are effectively identical.  The ICEM coefficient is
adjusted to these data since no other data are available to determine the
proper coefficient. The agreement of the calculation with the shape of the
distribution is excellent.  Applying this same
coefficient, along with the upper bound of $P_{{\rm icc}, 7}^0$, gives the
dot-dashed curve appearing in the bottom right corner.  (Note that only the
results from the LHCb Collaboration are shown here because the intrinsic charm
contribution to $X(3872)$ production measured at more central rapidities by
the CMS \cite{CMS_XofpT} and ATLAS \cite{ATLAS_XofpT} would not be visible on
the plot.)

This result shows that at
$\sqrt{s} = 13$~TeV, the contribution from intrinsic charm to the production of
$X(3872)$ is negligible.  This is not surprising given that average rapidity
for the $X(3872)$ at this energy is 7.42, see
Table~\ref{ave_rapidity_table}.  The boost is such that, in the rapidity
range given in Table~\ref{ave_pT_table}, $2.5 < y < 5$, only 0.15\% of
$P_{{\rm ic}, n}$ is contained in that rapidity interval.  Applying the interval
measured by LHCb, $2 < y < 4.5$, would reduce that value still further.  The
average $p_T$ in the given range is very high,
$\langle p_T \rangle \sim 48$~GeV.

\begin{figure}
  \begin{center}
    \includegraphics[width=0.495\textwidth]{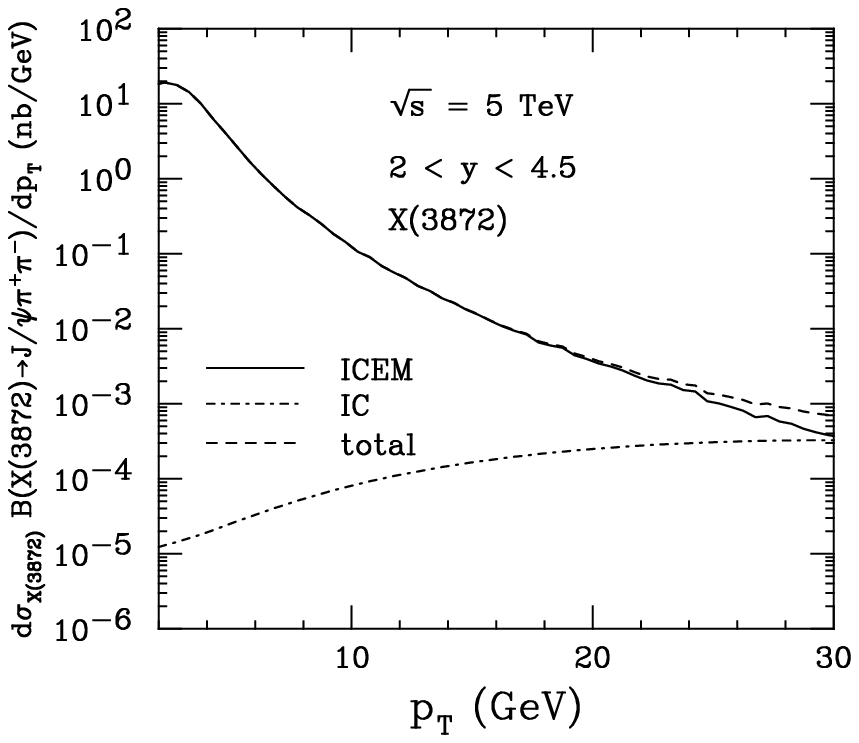}
  \end{center}
  \caption[]{The cross section of $X(3872)$ production as a function of
    $p_T$ at $\sqrt{s} = 5$~TeV and $2<y< 4.5$.
    The distribution calculated in the ICEM is shown by the solid line.
    The dot-dashed line gives the contribution from intrinsic charm in the
    rapidity range at this energy.  The sum of the two is given by the dashed
    line.
  }
\label{5TeV_XofpT}
\end{figure}

However, this does not rule out possible observation of tetraquark
production
by intrinsic charm at the LHC.  The contribution could be more than a factor of
10 higher at $\sqrt{s} = 5$~TeV, with a peak at significantly lower $p_T$, see
Table~\ref{ave_pT_table} and Fig.~\ref{ic_X_pTdists}.  While the intrinsic
charm contribution becomes larger at lower energies, due to the decreased
boost in rapidity, the perturbative production in an approach such as that of
the ICEM, decreases so that the two may become more competitive, allowing the
intrinsic charm contribution to become visible at high $p_T$, see
Fig.~\ref{5TeV_XofpT}.

\section{Summary}
\label{Summary}

The tetraquark mass and kinematic distributions have been studied in terms of
the intrinsic charm model.  The results suggest that a narrow $k_T$ range,
suggestive of tightly-bound partons is
compatible with most of the measured tetraquark candidate masses.  The mass
distributions also suggest that, for tetraquark candidates with one or two charm
quark constituents, the $X(3872)$, the $X_s$, and the $T_{cc}^-$ are compatible
with a meson pair structure for the tetraquark while, on the other hand, the
$T_{c \overline c s}$, the $T_{cs}^0$ and the $T_{cs0}^a$ are more compatible with a
loosely bound four-quark configuration.

The kinematic distributions calculated here, are assumed to
be independent of the structure.  At LHC energies, as studied here, the rapidity
distributions are boosted to high rapidity while the $p_T$ distributions are
very hard, with a high $p_T$ tail.  These kinematics are considerably different
than those obtained in perturbative QCD, as already noted for the $J/\psi$ and
$\overline D$ mesons \cite{RV_IC_EN}.

The potential cross sections in this approach are all small but could dominate
production in regions of kinematic phase space
where production by perturbative QCD
mechanisms is small, namely at higher rapidity and transverse momentum, as shown
in Ref.~\cite{RV_SMOG} for fixed-target $J/\psi$ and $\overline D$ production.
However, the cross sections given in Sec.~\ref{Rates} are total cross sections
and do not include any reduction due to finite detector acceptance which could
reduce them still further.  As shown in Figs.~\ref{X3872_LHCb} and
\ref{5TeV_XofpT}, the intrinsic
charm contribution to $X(3872)$ production is negligible at $\sqrt{s} = 13$~TeV
but could become more visible at 5~TeV where the distribution is less boosted.
If tetraquark candidates could be measured in either the fixed-target
environment of the LHCb SMOG device or at the future electron-ion collider, one
might see an even more significant effect due to intrinsic charm.

Finally, the same basic calculational structure can be applied to bottom
tetraquarks, as was already done for a potential
$X_b(b \overline b b \overline b)$ state in Ref.~\cite{ANDY}.   This will be
considered in future work. \\

{\bf Acknowledgments}
A.~Angerami and V.~Cheung are thanked for discussions.
This work was supported by the Office of Nuclear
Physics in the U.S. Department of Energy under Contract DE-AC52-07NA27344 and
the LLNL-LDRD Program under Project No. 23-LW-036, and the HEFTY Collaboration.

\end{document}